\documentclass[acmsmall]{acmart}

\usepackage{hyperref}
\usepackage{microtype}
\usepackage{amsmath}
\usepackage{mathtools}
\usepackage{amsthm}
\usepackage{cleveref}

\usepackage{graphicx}
\graphicspath{{figures/}}

\usepackage{ragged2e}
\usepackage{xcolor}
\usepackage{mdframed}
\usepackage{tabularx}
\usepackage{array}
\usepackage{booktabs}
\usepackage{multirow}
\usepackage{longtable}
\usepackage{soul}

\newcolumntype{L}[1]{>{\raggedright\arraybackslash}m{#1}}
\newcolumntype{R}[1]{>{\raggedleft\arraybackslash}p{#1}}
\newcolumntype{C}[1]{>{\centering\arraybackslash}p{#1}}

\setcopyright{cc}
\setcctype{by}
\acmJournal{PACMHCI}
\acmYear{2026} \acmVolume{10} \acmNumber{6} \acmArticle{CSCW123}
\acmMonth{10} \acmDOI{10.1145/3816971}

\begin{document}

\received{13 May 2025}
\received[revised]{13 January 2026}
\received[accepted]{17 March 2026}

\title[Method, Mind, and Morality]{Method, Mind, and Morality: How People Make Sense of Artificial Intelligence}

\author{Jacy Reese Anthis}
\affiliation{
  \institution{University of Chicago}
  \city{Chicago, IL}
  \country{USA}
}
\affiliation{
  \institution{Stanford University}
  \city{Stanford, CA}
  \country{USA}
}

\author{Erik Brynjolfsson}
\affiliation{
  \institution{Stanford University}
  \city{Stanford, CA}
  \country{USA}
}

\author{James Evans}
\affiliation{
  \institution{University of Chicago}
  \city{Chicago, IL}
  \country{USA}
}
\affiliation{
  \institution{Santa Fe Institute}
  \city{Santa Fe, NM}
  \country{USA}
}

\renewcommand{\shortauthors}{Jacy Reese Anthis, Erik Brynjolfsson, and James Evans}

\begin{abstract}
    How can humans make sense of the rapid takeoff of artificial intelligence (AI)? We studied the sensemaking dynamics of AI through an open-ended, mixed-methods study with computational text analysis of millions of AI-related newspaper articles and social media posts grounded in 57 semi-structured interviews with AI professionals in 2021 and 2023—before and after the recent surge of public interest. We identify a range of sociological frames (interpretive schemas that structure collective cognition) and show how AI professionals use frames to address significant cognitive challenges, such as assigning responsibility for societal impacts. We develop a framework of three primary debates across which frames are adopted and contested: (i) the \textit{method} of AI development, between frames of top-down expert systems and bottom-up emergent capabilities, (ii) the \textit{mind} of an AI system, ranging from a passive tool to a humanlike “digital mind,” and (iii) the \textit{morality} of how AI is used, particularly the decision of whether to slow down or speed up AI development. As humanity enters the era of transformative AI, technologists and policymakers must account for the framing dynamics that will circumscribe our beliefs, values, and actions.
\end{abstract}



\begin{teaserfigure}
    \centering
    \includegraphics[width=0.79\linewidth]{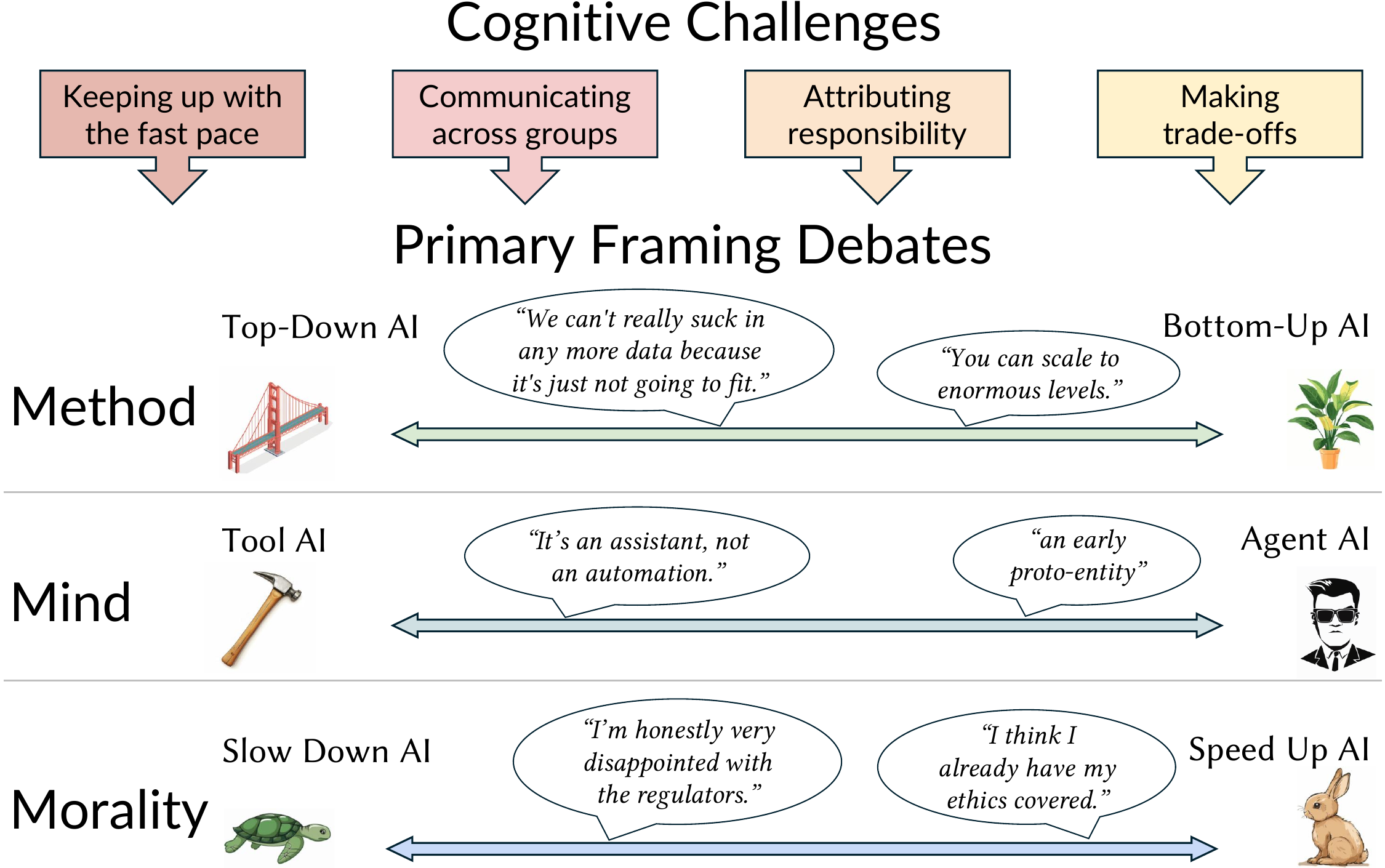}
    \caption{Summary of our conceptual framework. We identify four cognitive challenges faced by humans (\Cref{sec:challenges}) as they wrestle with three primary dimensions of what AI means for society (\Cref{sec:debates}).}
    \label{fig:conceptual_figure}
    \Description{Four boxes are shown at the top in shades of red and yellow, for each cognitive challenge. Arrows extend down to three rows for the three primary dimensions. Each row has the dimension label and a bidirectional arrow. Each end of each arrow has the label of the dimension, a small visualization (e.g., a bridge for ``top-down''), and a related participant quote.}
\end{teaserfigure}

\maketitle

\section{Introduction}

Artificial intelligence (AI) systems, particularly general-purpose large language models (LLMs), have rapidly permeated a wide variety of social contexts, including education~\citep{han_teachers_2024}, work~\citep{valentine_algorithm_2024}, politics~\citep{kobiella_if_2024}, and art~\citep{caramiaux_explorers_2022}. To adapt to these changes, people must incorporate AI into their existing worldviews and behavior. Within a single context, AI can have a variety of technical and social meanings: on social media, AI can mean the algorithmic feeds that deliver content from one user to another~\citep{jahanbakhsh_exploring_2023}, a simulated “bot” user that posts and amplifies content as if it were human~\citep{long_could_2017}, or AI applied to other contexts that is the subject of social media discussion~\citep{mao_online_2023}. These meanings differ between people: an AI-based algorithmic feed may be understood as an exciting feature for technologists and investors but as a threat to mental health for educators and parents.

In social computing, social meanings of AI span diverse contexts. At a high level, AI is sometimes viewed as a threat to human identity and agency~\citep{chen_trauma-informed_2022, milton_i_2023, register_attached_2023, sturgeon_humanagencybench_2025}, a “social actor” attributed moral agency and moral standing to do harm and be harmed~\citep{anthis_moral_2021, harris_moral_2021, lima_collecting_2020, reeves_media_1996}, and an economic force with the dual-use potential to replace or empower workers~\citep{li_user_2024, woodruff_how_2024}. Various mental models have been proposed, such as “role play” as AI systems rotate through diverse personas~\citep{shanahan_role_2023}. The number and diversity of social meanings continue to expand as entrepreneurs, journalists, and others use the term “AI” to command attention for their work. This broad usage may reduce the extent to which the term conveys meaning, but it nonetheless continues to be described as “transformative”~\citep{the_new_york_times_i_2024}, “a national priority”~\citep{lee_what_2018}, and “a general purpose technology”~\citep{eloundou_gpts_2023}. Social meaning has long been a central interest in psychology~\citep{frankl_mans_1946} and sociology~\citep{goldberg_sociology_2024}, and it is an increasingly common topic in human-computer interaction~\citep{mekler_framework_2019}.

This raises a provocation for social science and cognitive science: \textbf{How can people make sense of AI in light of this multitude of social meanings?} Past work in CSCW has examined particular social meanings, such as data scientists viewing AutoML as an augmentation of their productivity despite the apparent possibility of displacement~\citep{wang_human-ai_2019} or organizational decision-making, in which AI algorithms are able to supersede the “arbitrarily segmented decision spaces” and the “entrenched structures” of organizational divisions based on human capabilities~\citep{valentine_algorithm_2024}. These particular social meanings have been studied individually, but there is not yet a framework for how biological human beings can navigate the immense challenge of staying cognitively afloat during the turbulent rise of “digital minds”~\citep{anthis_perceptions_2025}.

To address this question, we conducted an open-ended, mixed-methods study with data from 2018 to 2024. Our empirical scope is deliberately broad, spanning the huge variety of AI domains rather than a single area (e.g., ethics, job automation) in order to focus on holistic sensemaking processes. In order to anchor meaningful findings in this massive and burgeoning field, we took advantage of recently developed methodologies that integrate computational “big data” and qualitative “small data” analysis into a single theory-building research pipeline~\citep{baumer_comparing_2017, berente_data-driven_2019, muller_machine_2016, nelson_computational_2020}.

Computationally, we constructed and analyzed topic models of AI-related newspaper articles (e.g., \textit{The New York Times}) and social media posts from Twitter (now known as X). Rather than conventional bag-of-words techniques such as latent Dirichlet allocation~\citep{blei_latent_2003} that leave out the semantic information conveyed by word order, we constructed topic models via dictionary learning across vector representations of words embedded in a high-dimensional space~\citep{arora_linear_2018, mikolov_efficient_2013}. Qualitatively, we analyzed the individual texts most representative of these topics, and we conducted 57 semi-structured interviews with AI professionals in two waves, 2021 (30 interviews) and 2023 (27 interviews). We asked participants about their experiences with the topics, their perceptions of topics, how topics relate to each other, and the co-evolution of topics over time.\footnote{Because of the iterative nature of our research process, the topics used for interviews in 2021 and 2023 were drawn from early versions of the computational models, built with only the documents archived before the interview was conducted (\Cref{fig:methodological_timeline}).}

The topic models provided a structured method to generate a list of topics for querying interview participants as well as a topography of the field on which to anchor their responses. After interviews and interview coding, the topic models contribute a latticework of frames and their relationships, while the interviews detail the dynamic processes by which individuals make sense of those frames. In line with the widely used paradigm of grounded theory~\citep{charmaz_constructing_2006, glaser_discovery_1967}, our approach built theory foremost from the empirical data rather than seeking the confirmation of preexisting theory or hypotheses. Our iterative research process, shown in \Cref{fig:methodological_timeline}, was built for this mode of inquiry rather than statistical hypothesis testing.

As we iterated between data analysis, theory development, and literature review, we situated our findings with the sociological theory of “frames,” sociocognitive schemas of interpretation by which people make sense of the world and take action~\citep{benford_framing_2000, goffman_frame_1974, norman_design_1988}. Computational social scientists have identified a congruence between framing theory and topic models~\citep{bail_cultural_2014, dimaggio_exploiting_2013, fligstein_seeing_2017, giorgi_many_2015, mohr_introductiontopic_2013}, and framing theory has precedent in human-computer interaction and social computing research, primarily to describe the online dynamics of social movements \cite[e.g.,][]{caldeira_how_2023, qin_dismantling_2024, stewart_drawing_2017, wu_negotiating_2024}. During unsettled times, such as technological emergence, frames are known to co-evolve through “framing contests,” in which social actors—including professionals, users, and journalists—advocate for their favored frames in an organization~\citep{kaplan_framing_2008}, social movement~\citep{ryan_prime_1991}, or field~\citep{hiatt_farms_2019}.

Our analysis brings together perspectives from public discourse and technology professionals. The two text corpora capture public discourse and journalistic mediation at scale, while the interviews provide in-depth accounts from people involved in AI development and deployment. Framing scholars have emphasized that frames circumscribe understanding across media producers and consumers~\citep{gitlin_whole_1980} as well as the general public and social “elites”~\citep{druckman_limits_2001}. We therefore study frames as circulating across particular social groups and the general public, subject to reconfiguration across contexts and institutions \citep[e.g.,][]{entman_framing_1993, van_gorp_constructionist_2007}.

Drawing on the text analysis, interviews, and framing literature, we make the following contributions:

\begin{itemize}
    \item We identify four cognitive challenges that motivate the choice and use of particular AI frames: \textbf{keeping up with the fast pace} of new information; \textbf{communicating across groups}, such as colleagues, friends, and family; \textbf{attributing responsibility} for societal consequences; and \textbf{making trade-offs} between different priorities and goals.
    
    \item We identify three primary framing debates that structure collective cognition: the \textbf{method} by which AI is created, ranging from frames of top-down expert systems to bottom-up emergence through massive scaling of computational resources; the \textbf{mind}\footnote{While some researchers hesitate to describe computer systems as “minds,” there is an extensive literature documenting how humans increasingly perceive mind in computer systems, particularly AI \citep[e.g.,][]{anthis_perceptions_2025, bullock_public_2025, reeves_media_1996, scott_you_2023}.} of what AI is perceived to be, ranging from a conventional software tool to a humanlike coworker or companion; and the \textbf{morality} of how to use AI, particularly the decision of whether to slow down or speed up its development.
    
    \item We consider potential implications of this high-level framework for design, policymaking, and research. It poses problems for AI design and policy, such as the difficulty of promoting nuanced positions when people default to particular frames, but also opportunities, such as being able to readily incorporate useful mental models that people already have from outside the AI context. For researchers, social dynamics—such as the coevolution of frames—can be integrated into economic and material-based models of the trajectory of AI technology, such as the potential role of framing contests in the much-discussed “productivity paradox”~\citep{brynjolfsson_productivity_1993} in which general-purpose technologies have delayed economic impact.
    
    \item While each method we incorporated has precedent in prior work, this study constitutes an example of how computational and qualitative paradigms can be integrated in unique ways to address the unique circumstances that arise during open-ended, data-driven research. In particular, we show how the “bird's eye” view of computational models can be enriched with the “worm's eye” view of participants' lived experience to characterize social dynamics of fast-paced technological change.
\end{itemize}

\section{Human-AI interaction}

Our aim in this project is to provide a high-level account of meaning-making in the interface of AI and human society. We summarize the dynamics of human-AI interaction, of which meaning is made, in three parts: the motivations for interaction—of which productivity has been the primary focus—the contextual factors that shape interaction and its outcomes, and existing research on the mental models of AI and human-AI interaction to which this project contributes. This presents a complex and cognitively demanding landscape for professionals and everyday users to navigate.

\subsection{Productivity and other motivations}

There are a variety of goals people have in mind when adopting AI technology, which shape how they think and speak about AI. Research on human-AI productivity suggests that human-AI collaboration can improve task performance in areas such as science~\citep{koepnick_novo_2019, wang_human-ai_2019} and creative writing~\citep{clark_creative_2018}. In the context of LLMs, the evidence now includes AI-use experiments with professionals, such as tests of writing tasks modeled on management consulting~\citep{dellacqua_navigating_2023}, and these publications have suggested promising results for human-AI teams compared to the performance of an unassisted human. However, the effect of AI use on productivity appears to vary widely across human contexts and AI capabilities. \citet{vaccaro_when_2024} performed a meta-analysis of dozens of experiments that measured the performance of human, AI, and human-AI participants. The meta-analysis showed that being on a team with an AI increased performance relative to the human alone, but team performance still lagged behind AI performance. Despite the popularity of AI, Vaccaro et al. show that only a few studies have documented “strong synergy,” in which teams outperform the best of either the human or AI participants.

The findings of Vaccaro et al. support the existence of an AI “productivity paradox”~\citep{brynjolfsson_productivity_1993}. AI is widely lauded as a general purpose technology~\citep{bresnahan_general_1995}, but economic research on general purpose technologies has documented substantial delays between introduction, such as that of computers in the 1970s and 1980s, and dramatic productivity gains, and AI may be undergoing its own productivity paradox~\citep{alderucci_quantifying_2019, brynjolfsson_artificial_2019}. Proposed explanations of the productivity paradox have tended to focus on material dynamics, such as the production of complementary technologies and the mismeasurement of new forms of economic value~\citep{brynjolfsson_productivity_1993}, but there may also be bottlenecks in productivity that manifest in the dynamics of social perception, interaction, and meaning-making—dynamics that could be clarified via framing theory.

Beyond productivity, AI systems affect social outcomes such as mental health and learning. AI-powered mental health interventions, such as chatbots and virtual therapists, have demonstrated promise in reducing symptoms of depression and anxiety~\citep{fitzpatrick_delivering_2017}, reducing loneliness and suicidal thoughts~\citep{maples_loneliness_2024}, improving the accessibility of mental health support~\citep{graham_artificial_2019}, and providing continuous personalized care in cases when a human practitioner could not~\citep{vaidyam_chatbots_2019, luxton_artificial_2016}. The integration of AI in mental health care has opened new avenues for public benefit but also risks, such as dependency and the generation of inappropriate content when people expect professional responses as they would a human mental health professional~\citep{lalvani_moral_2022} or when people enter a “delusional spiral” of reinforced delusions with AI companions~\citep{moore_characterizing_2026}. AI in various forms has been widely integrated into educational systems for decades. Since 2011, intelligent tutoring systems have shown comparable effects to human tutors in isolated experiments~\citep{vanlehn_relative_2011}, though others in education have pushed for the use of AI in data-driven design and data mining to empower human educators~\citep{baker_stupid_2016}. These motivations have led to a wide variety of social groups developing AI, deploying AI, and advocating for their interests in how AI is used in society.

\subsection{Contextual factors}

As people adapt to AI in new and diverse contexts, they need to make sense of the contextual factors that determine its effects. Task performance depends on the user's trust, which in turn depends on factors such as the user's age, culture, gender, and personality~\citep{hoff_trust_2015}. It depends on perceived intelligence and competence~\citep{zhang_effect_2020} and the transparency and explainability of the system~\citep{miller_explanation_2019}. User characteristics are consequential, including the user's AI literacy~\citep{long_what_2020, mughari_effect_2024, pinski_ai_2024}, cultural background~\citep{rau_effects_2009}, democratic and autocratic decision-making styles~\citep{munyaka_decision_2023}, and their perceptions of artificiality in the AI teammate~\citep{schelble_investigating_2023}. The design of the user interface that connects the user and system can also make a significant difference~\citep{amershi_guidelines_2019, strohmann_designing_2019}. Researchers have explored the design space of future AI tools, particularly with systems that augment the human writing process in either natural language~\citep{lin_rambler_2024, yuan_wordcraft_2022} or computer code~\citep{angert_spellburst_2023, mozannar_reading_2024}.

Human-AI interaction is often studied in salient contexts such as LLMs and robots, but, arguably, most human-AI interaction occurs with the more embedded AI systems that power online technologies such as online search and shopping~\citep{gaur_ai_2024}. This can result in complex interaction dynamics. As people browse, online behavior creates vast amounts of residual data, sometimes referred to as data exhaust, that is used to train future AI systems. Repeated feedback loops of interaction and system development lead to the amplification of bias~\citep{taori_data_2023}, surveillance capitalism~\citep{zuboff_big_2015}, and predictive policing~\citep{brayne_big_2017, brayne_predict_2021}. These co-evolving social dynamics pose difficulties for sensemaking and taking meaningful social action to steer the trajectory of AI.

\subsection{Mental models}

Cognitive scientists and computer scientists have documented a variety of mental models that people use to make sense of computers and AI. A long-observed dynamic is that humans readily attribute humanlike features to AI. In the 1960s, the chatbot ELIZA was able to dazzle users with humanlike “talk therapy” conversation by using simple heuristics. Cognitive scientist Douglas Hofstadter called the tendency to see mental faculties in even simple computers the “ELIZA effect”~\citep{hofstadter_go_1979}. We have long known that “computers are social actors”~\citep{nass_machines_2000, nass_computers_1994, reeves_media_1996} and that humans tend to anthropomorphize AI~\citep{epley_seeing_2007, melson_robots_2005, watson_rhetoric_2019, zlotowski_anthropomorphism_2015}, particularly via the attribution of mental states, such as agency and experience~\citep{gray_dimensions_2007, kozak_what_2006, lima_human_2021, scott_you_2023}. Recent experimental research has identified mental models of autonomy and sentience that undergird AI perceptions~\citep{pauketat_mental_2026}.

Studies have applied concepts from philosophy and moral psychology, developed in the exclusively human context, to human-AI interaction. They find that users readily attribute moral agency and responsibility~\citep{komatsu_blaming_2021, manoli_ai_2025, stuart_guilty_2021} as well as concern for the moral standing of AI entities, such as their rights and welfare~\citep{ladak_extending_2023, ladak_substratism_2026, ladak_which_2024, lima_collecting_2020}. Recent survey-based research has provided quantitative assessments of how everyday people perceive mind and morality across a variety of AI systems~\citep{ladak_robots_2025}; they show that AI systems are perceived to have mind and moral status within a relatively narrow band in the scope of a broad range of entities (e.g., rocks, humans) but that there is significant variation within that. 

HCI researchers and designers have developed various taxonomies of social meaning. In terms of work, the AI can be a tool (low AI autonomy and low human involvement), servant (low AI autonomy and high human involvement), assistant (high AI autonomy and low human involvement), or mediator (high AI autonomy and high human involvement)~\citep{kim_one_2023}. In terms of communication, the AI can be a creator (communicating with many humans), curator (mediating communication with many humans), converser (communicating with one human), or co-author (mediating communication with one human)~\citep{sundar_rethinking_2022}. Other frameworks typically contrast tool-like and agent-like roles~\citep{chen_portraying_2025, dautenhahn_what_2005, takayama_beyond_2008, xu_tool_2024}. In some studies, the AI is presented to the user with a particular metaphor chosen by the researchers, such as testing the effects of describing a chatbot as a book, dog, toddler, or trained professional, among other metaphors, on users' perceived warmth and competence~\citep{jung_great_2022, khadpe_conceptual_2020}. Recent work has emphasized the importance of understanding and shaping the social roles filled by AI~\citep{earp_relational_2025}, and a rapidly emerging literature attempts to scientifically analyze the novel “digital companionship” with AI systems~\citep{manoli_digital_2026} and understand how AI companionship develops~\citep{hwang_how_2025}.

Other frameworks categorize dynamic features of the human-AI relationship, such as dispositional, situational, and learned trust~\citep{hoff_trust_2015}. These dynamic features are increasingly viewed as multidimensional, such as autonomy expanding from the conventional single dimension between more human involvement and more AI involvement~\citep{beer_toward_2014} to—given the human and AI can both be more or less involved—disentangling operational, intentional, shared, non-deterministic, physical, and cognitive autonomy~\citep{kim_taxonomy_2024}.

While these mental models generalize across different contexts, the frameworks have tended to focus on particular elements of human attitudes and AI behavior, rather than the ways in which social groups manage the cognitive demands of AI technology. Nonetheless, these mental models can provide a connective tissue from individual interaction to macroscopic theories of AI and society, such as algorithmic fairness~\citep{anthis_impossibility_2024}, algorithmic work and control~\citep{kellogg_algorithms_2020}, and the posited economic transformation of a “second machine age”~\citep{brynjolfsson_second_2014}. In our study, these mental models of individual AIs and of AI technology as a whole are important constituents of sensemaking, framing, and the longitudinal process of psychological world-making~\citep{pauketat_world-making_2025}.

\subsection{AI discourse}

The study of AI discourse at scale remains relatively small, in part because AI has only in recent years taken the public spotlight. Nonetheless, there have been a variety of studies. For example, \citet{fast_long-term_2017} analyzed 30 years of AI coverage in \textit{The New York Times}, mapping out keywords alongside sentiment measures and documenting both growing concern about losing control of AI and hope for beneficial impacts such as healthcare.

Other studies have developed or identified particular AI frames, discourses, and narratives. \citet{johnson_reframing_2017} argued for “A New Frame for AI Discourse”: viewing an AI system as including both computational artifacts and human actors, and cautioning that AI-risk debates can confuse machine autonomy for the human autonomy that drives technological harm. \citet{bearman_discourses_2023} reviewed references to AI in higher education journals and identified two “discourses”: AI as an “imperative change” and AI as “altering authority” to include not just teachers but other staff, students, and nonhuman entities such as machines and corporations. Recently, \citet{gilardi_we_2024} examined the volume and sentiment of AI news and proposed four narratives: existential risk, effective accelerationism, immediate societal risks, and balanced risks. While \citet{gilardi_we_2024} foreground important narratives, we integrate large-scale text analysis with interviews to map cross-cutting framing debates in which such narratives can emerge, change, and dissipate.

\section{Framing theory}

We began our study with a broad interest in how people make sense of AI, particularly the attribution of social meaning. However, the sociological study of meaning has been referred to as “elusive”~\citep{patterson_making_2014} and an “exciting but disjoint field”~\citep{goldberg_sociology_2024}. After initial data collection, we honed our interest to the sociological theory of “frames,” the schemas of interpretation that render meaning and organize experience~\citep{goffman_frame_1974}. Framing may be particularly useful to explain collective meaning-making and social change as AI spreads throughout society, better than related concepts such as individual interpretation, because framing “is a form of public culture, while interpretation is private”~\citep{goldberg_sociology_2024}.

Frames also cut across social groups and institutional roles: for example, \citet{gitlin_whole_1980} argues that media frames organize the world both for journalists and for audiences who rely on their reports, and \citet{druckman_limits_2001} notes that public opinion often depends on which frames elites choose to use. More broadly, framing scholars emphasize that framing links communicators and receivers across institutional sites \citep[e.g.,][]{entman_framing_1993, van_gorp_constructionist_2007}. In studying AI meaning-making, we therefore attend to frames circulating between AI professionals and public discourse—through conventional media, social media, and lived experiences—even as they are reconfigured across contexts and institutions.

From the expansive framing theory literature, we draw on framing dynamics and categories of frames particularly relevant to AI meaning-making: dynamics such as alignment through which frames are sculpted and selected by individual cognition, the strategic and technological frames that are particularly relevant for AI, the notion of discursive opportunity structures by which frames can circumscribe future action, and the notion of framing contests by which frames are deliberately pitted against each other to further individual and collective interests.

\subsection{Sensemaking, resonance, and alignment}

Scholars of social movements, organizations, technology, and other domains have developed a rich vocabulary to describe frames and the framing processes by which they coevolve over time. Within a single individual, meaning-making is often referred to as \textit{sensemaking}~\citep{weick_collapse_1993}, which emphasizes the causal process by which an individual makes sense of their material environment. The concept was developed through the study of the Mann Gulch fire of 1949 in which firefighters struggled to make sense of a new situation~\citep{weick_sensemaking_1995}. The firefighters who perished remained fixated on their training, holding onto their equipment and attempting conventional techniques, while the foreman who survived was able to make sense of the new situation: dropping his tools and setting a fire to create a safe haven while the larger fire passed him by~\citep{weick_sensemaking_1995}.

In terms of sensemaking, framing can be conceptualized as sense-giving, the collective process by which frames are not only developed but modified and contested by social actors~\citep{fiss_discourse_2005}. Once a new frame has come into existence, either through individual experience or through past framing processes, its reach and longevity depend on its \textit{resonance}, the extent to which the frame accords with the interests of particular individuals or groups, such as its compatibility with one's everyday experiences or cultural heritage~\citep{snow_ideology_1988}. Resonances allow social actors to shape the \textit{alignment} between frames and the interests of a particular social group through processes such as bridging multiple frames together~\citep{snow_frame_1986}.

Resonance and alignment have been used to explain a variety of social movement phenomena, such as what drives movement participation: Why do different movements draw vastly different levels of engagement? Why does participation rise and fall over time? When people do participate, why does that participation take many different forms? Each type of variation can be, at least in part, explained by frame alignment. For example, groups tend to participate most when the social movement's grievances are aligned with that group's interests~\citep{snow_frame_1986}.

\subsection{Strategic and technological frames}

Frames can be categorized into a variety of types, and the frames most relevant for our study of AI are strategic frames and technological frames—concepts primarily developed in organizational sociology and management studies. Modern framing theory builds on a long history of related concepts in these fields, such as the psychological notion of a “frame of reference” through which individuals interpret information, such as making that information validate their preconceptions \cite[e.g.,][]{march_organizations_1958, starbuck_organizations_1983}, as well as “scripts” and Marvin Minsky's notion of “frames” as computer science data structures~\citep{gioia_scripts_1984}. 

Management scholars describe a “strategic frame” as one deliberately adopted by an entire firm or industry to advance their interests. For example, \citet{huff_industry_1982} discusses how Volkswagen succeeded with the strategic frame of being the “people's car” until 1964, when they lost track of that frame and lost cultural relevance until, in 1971, the company prioritized a new singular frame, even removing many of its existing car models to ensure its focus \cite[for details, see][]{mintzberg_patterns_1978}. \citet{lounsbury_cultural_2001} articulated the idea of a “field frame,” a strategic frame used for the status and legitimacy of an “organizational field”~\citep{dimaggio_iron_1983}. In their case study, Lounsbury et al. detail how the recycling industry in the postwar decades fought against the monolithic “resource recovery” frame, which largely equated recycling and the incineration of garbage—obfuscating the advantages of recycling. By shifting discourse towards “resource recovery and recycling,” proponents were able to make the advantages more salient, leading to the rapid expansion of the recycling industry.

While most of the framing literature has focused on social movements and organizations, scholars have developed a number of related concepts to explain evolving interpretations of science and technology. The most well-known model, the punctuated equilibrium of “paradigm shifts”~\citep{kuhn_structure_1962}, has explained a variety of scientific advances, such as from Newtonian to quantum mechanics. However, more recent scholarship has foregrounded social construction~\citep{pinch_social_1984}—in contrast to techno-determinism in which scientific research is a force \textit{causa sui} advancing monotonically towards scientific truth. Science and technology are also determined by human values, identities, norms, and other social phenomena that constitute frames, with \citet{orlikowski_technological_1994} explicating the notion of a Goffmanian “technological frame” in 1994. They also highlighted the overlap with a number of closely related terms for “shared cognitive structures” in this evolving interdisciplinary literature, such as “schema,” “paradigms,” and those more well-established in CSCW: “mental models” and “scripts.” In their case study, they drew out the wide gulf between technologist and user frames: technologists tend to consider technology in isolation from its social context—evaluating a new artifact solely based on technical metrics, while users think of technology as embedded in their everyday environments in the context of existing practices and behaviors. This gulf can be perilous as technologists often fail to account for such differences and products fail to make the transition from design to commercialization.

\subsection{Discursive opportunity structures}

Framing theorists have posited “discursive opportunity structures” as the social phenomena “within the symbolic realm” that determine which ideas are sensible or realistic for a certain group at a certain time~\citep{koopmans_political_1999}. These structures are juxtaposed with “political opportunity structures,” which entail a “narrow institutional sense” of opportunity. For example, a group of young people could lack the enfranchisement to directly influence electoral politics, but they could still utilize discursive opportunity structures, such as the ability to post content on social media or organize public demonstrations.

In the context of an emerging technology, structural opportunity is particularly important because frames are often not yet in place. Frames can be drawn directly from adjacent fields, such as existing related technologies, but they can also be drawn from more abstracted “master frames,” such as a moral notion of “fairness” or “justice” on which people build domain-specific frames. Master frames are particularly important in these rapidly growing fields, including AI, because there is less preexisting discursive opportunity structure on which to build.

\subsection{Framing contests}

Different frames, structures, and processes interact via overarching “framing contests,” a term coined by \citet{ryan_prime_1991} in the context of political activism to explain the “battleground” of culture. \citet{ryan_prime_1991} conceptualized framing contests through the strategies activists and their adversaries can use, both attacks (e.g., “discrediting via extreme cases”) and defenses (e.g., “avoidance of restatement”). Other framing theories have also found traction in such descriptions. For example, \citet{gamson_media_1992} argued that, in the context of media, frames are “indispensable and elusive,” and media evolves via “arenas” and “contests” of meaning-making.

The most well-known model of framing contests comes from \citet{kaplan_framing_2008} as a model for contestation within a single organization, in which individuals and within-organization groups use frames as repertoires or “tool kits”~\citep{swidler_culture_1986} to shape the strategic frames of the organization. Kaplan's model can be summarized in three stages. First, actors put forth various frames that each have particular resonances with other actors. Second, actors engage in social interaction that contests the legitimacy of frames and their proponents and contests the alignment of those frames with other actors. Third, based on the resulting resonance, the organization adopts a predominant collective frame—typically a strategic frame for the organization as a whole—or continues with divergent frames, deferring organizational decisions until after further contestation.

In more recent work, framing contest theories have been partially expanded to the level of fields, such as diesel particulate filters~\citep{guerard_turning_2013}, pay TV~\citep{gurses_entrepreneurship_2015}, and biofuels~\citep{hiatt_farms_2019}. Theoretical contributions of those studies include the idea of thresholds that frames must reach in a paradigm-like punctuated equilibrium, the ability of organizations to create turmoil that forces change even in highly regulated industries, and the extension of discrete frames to a continuous dimension of meaning in which actors push for increases or decreases rather than categorically new meanings.

In CSCW, meaning-making has largely been explored with individuals or small groups digesting information from the outside world \cite[e.g.,][]{calota_assembling_2024} or in the frames and counterframes built and deployed by opposing factions in an online debate \cite[e.g.,][]{stewart_drawing_2017}. While the notion of multi-polar meaning-making is not unknown to CSCW researchers, such as in the divergent social media “realities” built by influencers~\citep{beers_influencer_2023}, there is room for enrichment by mapping the topography of framing debates across multi-polar, highly contested fields. Indeed, with the globalization of discourse and entwinement of digital conflicts, this is arguably the primary mode in which modern meaning-making actually occurs.

We do not aim to directly advance framing theory outside of the AI context, but to use this literature to develop conceptual infrastructure for AI meaning-making. In contrast to much of the AI ethics and philosophy literature—which primarily advances principles and recommendations for how AI should be developed and governed \citep[e.g.,][]{floridi_ai4peopleethical_2018, jobin_global_2019, fjeld_principled_2020, mittelstadt_principles_2019}—our contribution is primarily descriptive: we characterize the frames being contested and the dynamics through which they are advocated and resisted. We also aim to be synthetic rather than centered on a single debate: for example, prior work has debated method (e.g., the “bitter lesson”~\citep{sutton_bitter_2019} and tensions between bottom-up and top-down approaches \citep[e.g.,][]{bender_dangers_2021, wei_emergent_2022}), mind (e.g., tool versus agent \citep[e.g.,][]{gray_dimensions_2007, reeves_media_1996, scott_you_2023}), or morality (e.g., arguments for slowing down AI \citep[e.g.,][]{hao_empire_2025}). Our contribution is to zoom out from these individual debates and provide evidence of how these ideas ebb and flow as actors bridge, contest, and align social meanings.

\section{Methodology}
\label{sec:methodology}

Our mixed-methods research process iterated between text analysis, interviews, literature review, and theory development. Rather than using our empirical methods to validate theory, as is typical in deductive and hypothesis-driven studies, we began with the data and followed the theoretical direction it admitted. This is aligned with \citet{nelson_computational_2020}'s “computational grounded theory” methodology and a host of similar innovative methodologies that have been developed over the past ten years~\citep{baumer_comparing_2017, berente_data-driven_2019, brayne_predict_2021, hannigan_topic_2019, muller_machine_2016, nelson_computational_2020, saetre_generating_2021}. These mixed-methods approaches build on constructivist and classical grounded theory methodologies that historically have relied on qualitative data~\citep{charmaz_constructing_2006, glaser_discovery_1967}, though neither these approaches nor our study strictly mimics a particular grounded theory methodology.

By integrating disparate viewpoints across the methodologies (text analysis and interviews), datasets (e.g., conventional and social media), and individuals (e.g., authors, participants), we are able to document a variety of social structures and processes present in the field of AI. This facilitated the theoretical depth that supported our data-driven theory development. This allowed us to interleave data collection with revisiting the academic literature, developing a framework that is primarily driven by empirical data but does not neglect prior work.

\begin{figure}[htbp]
    \centering
    \setlength{\fboxrule}{1pt}
    \setlength{\fboxsep}{4pt} 
    \includegraphics[width=\linewidth]{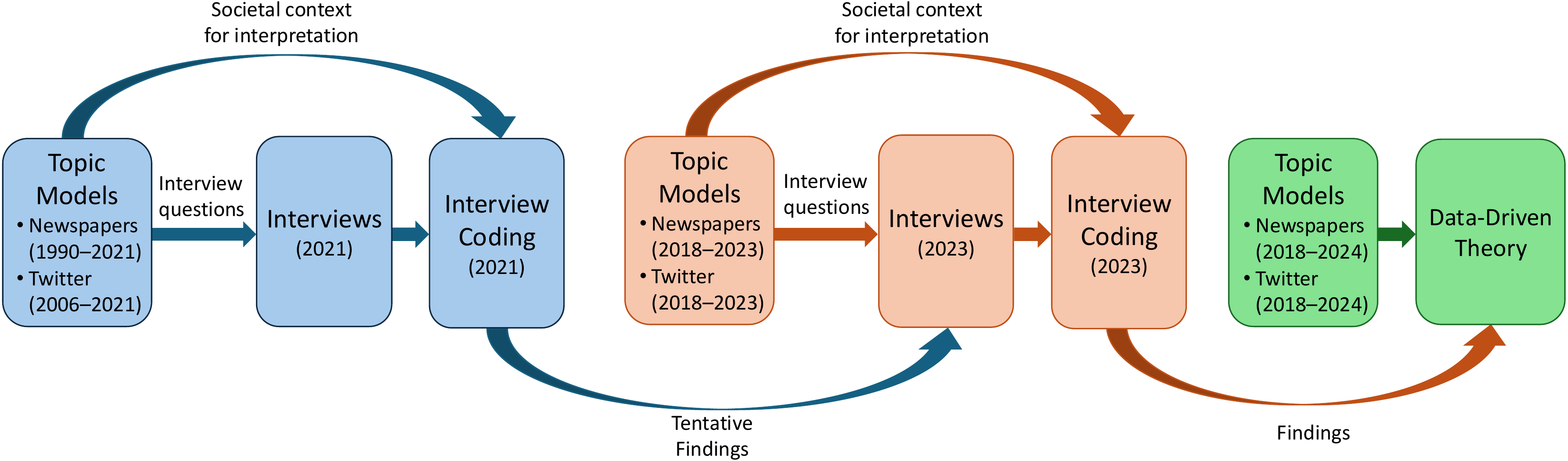}
    \caption{The project developed data-driven theory by iterating between computational modeling and semi-structured interviews with professionals in the field of AI. Events between the first and second set of interviews that informed theory development include the releases of ChatGPT, Gemini, and Claude.}
    \label{fig:methodological_timeline}
    \Description{The timeline is made of colored boxes and arrows, representing three sets of topic models (2021, 2023, and after the 2023 interviews). The topic models for 2021 and 2023 are shown as providing the interview questions for that year.}
\end{figure}

The focus of the study was the period from the first round of interviews in mid-2021 to the second round of interviews in mid-2023 when several state-of-the-art AI systems were released and popularized, including AlphaFold 2 in July 2021~\citep{jumper_highly_2021}, InstructGPT in January 2022~\citep{ouyang_training_2022}, DALL-E 2 in April 2022~\citep{ramesh_hierarchical_2022}, OpenAI ChatGPT in November 2022~\citep{openai_introducing_2022}, Google Gemini in February 2023~\citep{pichai_important_2023}, and Anthropic Claude in March 2023~\citep{anthropic_introducing_2023}. We included earlier texts from the corpora, dating back to 2018, and later texts, dating up to mid-2024, for additional social context. A summary of the project timeline is provided in \Cref{fig:methodological_timeline}.

\subsection{Corpora}

We selected two text corpora to document discourse in traditional media (newspaper articles) and social media (Verified posts or “tweets” from Twitter, which is now officially known as X). These corpora were selected based on public accessibility and tracking significant domains of social discourse, though each has significant limitations, such as being authored by non-representative subsets of the human population (\Cref{sec:limitations}).

In both corpora, we searched for documents containing at least one term closely related to the term AI: artificial agent, artificial assistant, artificial intelligence, autonomous vehicle, big data, bot, chatbot, and others listed in \Cref{tab:search_terms}. Beyond these search terms, we did not apply contextual filters to restrict topics or viewpoints, instead aiming to cast a wide net over public AI discourse.

For newspaper articles, we used an extensive global collection of English newspapers from ProQuest, such as \textit{The Times}, \textit{China Daily}, \textit{The New York Times}, \textit{The Wall Street Journal}, \textit{The Times of India}, and \textit{The Irish Times}. This resulted in 530,445 articles ranging from January 1, 2018 to June 30, 2024—excluding earlier articles to provide semantic information more closely associated with the period of change spanned by our 2021 and 2023 interviews. Filtering out non-English texts, exceptionally short texts (e.g., only a headline), and low-quality texts based on the frequency of typographical errors (e.g., due to poor optical character recognition)—targeting the exclusion of 30\% of articles—we had a final corpus of 371,312 articles with an average of 789 words. Because of the long-term nature of this study, with data collection beginning in 2021, we refreshed the computational models as the study progressed. Additionally, because newspaper articles often cover several domains, we only analyzed the document text that was within two sentences of a search term, extracting an average of 84 words. For sensitivity analysis, we also built models with the complete newspaper articles, which led to similar results except for the presence of fewer AI-specific topics.

For tweets, we used the Twitter academic API to gather tweets from January 1, 2018 to April 20, 2023 (the latest date available before the Twitter API was shut down later in 2023). Because of the scale of the platform and constraints of the Twitter API as well as the quantity of low-quality content, we treated Verified accounts as the primary social media corpus and, within this, aimed to cast a wide net. Tweets from Verified accounts more consistently represented genuine discursive contributions from real people and organizations, but we acknowledge the limitation that it is not representative of all social media discourse. We gathered 1,391,195 tweets with an average of 24.0 words each. For sensitivity, we also built models with non-verified tweets for comparison, finding similar results with more noise in the models.

In order, we pre-processed all text by removing URLs, numbers, punctuation, symbols, emojis, and stop words (including corpus-specific terms such as “amp” and “rt” on Twitter), lower casing the text, phrasing frequent bigrams (e.g., “big\_data”), and lemmatizing verbs (e.g., “kicked” to “kick”).

\subsection{Discourse atoms}

Conventional topic models such as latent Dirichlet allocation (LDA) view each document as a “bag of words,” a collection of words without any internal order or structure~\citep{blei_latent_2003}. Because of the breadth of our analysis across the field of AI, we used a more recently developed method based on word embeddings, representations of words via high-dimensional vectors, typically in a space with hundreds of dimensions. This kind of mathematical representation is the foundation of large language models such as ChatGPT, and as a research tool, they have been used with impressive results in the social sciences to study topics such as culture~\citep{kozlowski_geometry_2019, stoltz_cultural_2021}, knowledge~\citep{shi_science_2020, sourati_accelerating_2021}, and criminology~\citep{arseniev-koehler_integrating_2022}.

To create topic models from word embeddings, we use the “discourse atoms” method that identifies discursive building blocks that can be used to construct the word vectors themselves~\citep{arora_linear_2018}. Through k-means clustering and singular value decomposition of the embeddings, discourse atoms are a set of vectors in the same space as the word embeddings. Each discourse atom has a set of nearest word vectors that can be viewed as a summary of the topic. In addition to capturing more semantic information than LDA and other conventional topic models, many more topics can be produced without losing coherence. While LDA models typically lose coherence as the number of topics is raised to ten to 20, discourse atoms can typically be scaled to 100 or 200 topics before losing coherence.

\subsection{Interviews}

\setlength{\tabcolsep}{4pt}
\begin{table}[hb]
    \caption{Summary of participants. Most respondents who identified as a manager or entrepreneur had prior experience in another listed occupation (e.g., data scientist).}
    \label{tab:participants_summary}
    \footnotesize
    \begin{tabular}{R{0.19\linewidth} C{0.03\linewidth} | R{0.16\linewidth} C{0.09\linewidth} C{0.09\linewidth} | R{0.17\linewidth} C{0.13\linewidth}}
        \toprule
        \multicolumn{1}{c}{\textbf{\multirow{1}{*}{\shortstack{}}}} &
        \multicolumn{1}{c}{\textbf{\multirow{1}{*}{\shortstack{}}}} &
        \multicolumn{1}{c}{\textbf{\multirow{1}{*}{\shortstack{Country or \\ Region}}}} &
        \multicolumn{1}{c}{\textbf{\multirow{1}{*}{\shortstack{\% Current \\ Location}}}} &
        \multicolumn{1}{c}{\textbf{\multirow{1}{*}{\shortstack{\% Original \\ Location}}}} &
        \multicolumn{1}{c}{\textbf{\multirow{1}{*}{\shortstack{Occupation}}}} &
        \multicolumn{1}{c}{\textbf{\multirow{1}{*}{\shortstack{\% Occupation}}}} \\[3.5ex]
        \toprule
        \textbf{Male (\%)} & 77\% & \textbf{US} & 50\% & 30\% & \textbf{Manager} & 20\% \\
        \textbf{Age (mean)} & 41 & \textbf{India} & 14\% & 23\% & \textbf{Data scientist} & 18\% \\
        \textbf{Years in AI (mean)} & 9 & \textbf{UK} & 9\% & 9\% & \textbf{Researcher} & 16\% \\
        \textbf{Years in AI (median)} & 5 & \textbf{South America} & 5\% & 5\% & \textbf{Software engineer} & 14\% \\
         &  & \textbf{Eastern Europe} & 2\% & 5\% & \textbf{Entrepreneur} & 9\% \\
         &  & \textbf{Middle East} & 2\% & 5\% & \textbf{Consultant} & 7\% \\
         &  & \textbf{Australia} & 2\% & 2\% & \textbf{Advisor} & 2\% \\
         &  & \textbf{Belgium} & 2\% & 2\% & \textbf{Designer} & 2\% \\
         &  & \textbf{Canada} & 2\% & 2\% & \textbf{Diplomat} & 2\% \\
         &  & \textbf{Ireland} & 2\% & 2\% & \textbf{Lawyer} & 2\% \\
         &  & \textbf{Italy} & 2\% & 2\% & \textbf{Product manager} & 2\% \\
         &  & \textbf{Singapore} & 2\% & 2\% & \textbf{Professor} & 2\% \\
         &  & \textbf{Sri Lanka} & 2\% & 2\% &  \textbf{Salesperson} & 2\% \\
         &  & \textbf{Japan} & 2\% &  & & \\
         &  & \textbf{Central America} &  & 2\% &  &  \\
         &  & \textbf{China} &  & 2\% &  &  \\
         &  & \textbf{Germany} &  & 2\% &  &  \\
         &  & \textbf{West Africa} &  & 2\% &  &  \\
        \bottomrule
    \end{tabular}
\end{table}

We conducted 57 semi-structured interviews with AI practitioners in a procedure approved by the first author's institutional review board (IRB). Participant recruitment was conducted through LinkedIn. The first author searched for virtual LinkedIn events on the topic of AI in July 2021 because events provide public lists of attendees and indicate current interest. The profile of each attendee was checked with an invitation sent if it contained at least one reference to AI (e.g., “artificial intelligence,” “deep learning,” “natural language processing”). Of the 57 interviews, 30 were conducted in 2021, and 27 were conducted in 2023. The same professionals were invited each time, resulting in 13 participants interviewed in both periods and 14 who did not participate when first invited but then did when invited in 2023 (\Cref{tab:participants_individual}).

There was no financial compensation. While our goal was not to gather a representative sample, we note that all participants are English-speaking, and they were likely more experienced (e.g., managers) and more engaged in the AI community in online networking and events than the average AI practitioner, among other possible differences.

We provide summary participant demographics in \Cref{tab:participants_summary} and individual details in the appendix (\Cref{tab:participants_individual}). Excluding three participants who declined to answer the optional demographic questions, the mean age was 41 years old, and 77\% were male. Countries of residence included the U.S. (50\%), India (14\%), the U.K. (9\%), and eleven others. Countries of origin included the U.S. (30\%), India (23\%), the UK (9\%), and fourteen others. Occupations included manager (20\%), data scientist (18\%), researcher (16\%), and approximately ten other occupations. The mean years of experience in the field of AI was nine, ranging from zero to 40, though these estimates are also only approximations because of the various ways of scoping the field. Ages and years of experience are reported as of 2021. 

All but one interview were conducted by video call, lasting between 30 and 92 minutes. One interview was conducted in person at the request of the participant. We followed a semi-structured interview protocol. The first part of the interview (5–15 minutes) focused on documenting the participant's background. The rest of the interview primarily focused on participants' experience with and opinions on various AI frames drawn primarily from the topic models built prior to the interview phase, as shown in \Cref{fig:methodological_timeline}.\footnote{Because frames are not a well-known concept among the general public, we did not explicitly mention it to participants unless they asked for more information about the study.} In other words, the topic-model frames were used to structure the interviews, but participants were able to introduce their own frames and other perspectives beyond the prompts.

Interview audio was recorded, transcribed, and coded with open coding and axial coding by the first author, following the constant comparative method~\citep{glaser_discovery_1967}. While much interview analysis in CSCW involves coding by multiple researchers, single researcher coding is common in fields such as sociology, in which many projects involve over 50 interviews \cite[e.g.,][]{brayne_big_2017, douds_diversity_2021, knight_carceral_2024}, and we caution against requirements for inter-coder reliability that risk imposing positivist structure on interpretivist methodology~\citep{mcdonald_reliability_2019}. In the data collection and analysis, we aimed for theoretical saturation—in which further computational models, readings, and interviews were resulting in few if any new themes. In total, there were 49 hours of interview recordings with 841 first-order codes and 213 higher-order codes.

\section{Findings}
\label{sec:findings}

In this section, we first summarize the AI frames identified through the discourse atom topic models. Second, we present four challenges that we identified primarily from interviews, which participants used frames to address. Third, we articulate the three primary debates in which frames were contested.

\subsection{Frames}

We identified a multitude of atomic frames present in the field of AI via discourse atom topic models, a selection of which we show in \Cref{tab:frames} with measurement of frame prevalence over time in the appendix (\Cref{fig:frames_over_time}). Because of the wide variety of frames present in the discourse, we also manually generated four broad categories of frames: components of sociotechnical AI systems, contexts of AI application, dynamics of AI systems over time, and issues of AI and society. We found that, in general, the news corpus had relatively more representation of components (e.g., Smart Home devices), and the Twitter corpus had relatively more representation of issues (e.g., European law)—reflecting the overlapping yet distinct nature of conventional and social media discourse.

\onecolumn

\renewcommand{\arraystretch}{1}
\setlength{\tabcolsep}{2pt}
\begin{longtable}{L{0.23\linewidth} L{0.07\linewidth} L{0.7\linewidth}}
    \caption{A selection of the frames identified through k-SVD dictionary learning. Rather than raw vector form, each topic is presented with a researcher-generated label as a shorthand for its semantic content, the corpus it is drawn from, and its five nearest words from the semantic space (with a count of at least 100 in the corpus). Frames are loosely grouped into four broad categories: components, contexts, dynamics, and issues.}
    \label{tab:frames} \\
    
    \multicolumn{3}{c}{\textbf{\multirow{1}{*}{\shortstack{\large Components of sociotechnical AI systems}}}} \\[.1ex] \toprule
    \multicolumn{1}{c}{\textbf{\multirow{1}{*}{\shortstack{Topic Label}}}} &
    \multicolumn{1}{c}{\textbf{\multirow{1}{*}{\shortstack{Corpus}}}} &
    \multicolumn{1}{c}{\textbf{\multirow{1}{*}{\shortstack{Nearest Words}}}}  \\[.5ex]
    \hline
    Chips & News & chip chips processors implanted neuromorphic \\ \hline
    Companies & Twitter & companies organizations businesses leaders orgs \\ \hline
    Devices & Twitter & phones devices ipads smart\_speakers appliances \\ \hline
    GPT-4 & News & gpt4 gpt3 generative chatgpt dalle \\ \hline
    GPUs & Twitter & gpus inferencing nvidia\_gpus inference workloads \\ \hline
    Large Language Models & News & models model\_llm model models\_llms llms \\ \hline
    OpenAI & News & openai ousted altman meta openais \\ \hline
    Pattern Recognition & News & semantic analysis pattern\_recognition image\_analysis analyse \\ \hline
    Smart Home & News & smart\_speakers amazons\_alexa alexa voicecontrolled amazon\_echo \\ \hline
    Tech Leaders & Twitter & zuckerberg andrew\_yang sentience elon\_musk mark\_zuckerberg \\ \hline
    Venture Capital & News & venture\_capital ventures funds private\_equity fund \\ \hline
    & Twitter & venture\_capital sensetime venturecapital fund has\_invested \\ \hline
    Vision & News & images image hyperrealistic text\_images images\_and\_videos \\
    \bottomrule
    \\
    
    \multicolumn{3}{c}{\textbf{\multirow{1}{*}{\shortstack{\large Contexts of AI application}}}} \\[.5ex] \toprule
    \multicolumn{1}{c}{\textbf{\multirow{1}{*}{\shortstack{Topic Label}}}} &
    \multicolumn{1}{c}{\textbf{\multirow{1}{*}{\shortstack{Corpus}}}} &
    \multicolumn{1}{c}{\textbf{\multirow{1}{*}{\shortstack{Nearest Words}}}}  \\[.5ex]
    \hline
    Art & Twitter & music paintings artists turns converts \\ \hline
    Banking & Twitter & banking positively pages\_and\_creators jones\_artwork profound \\ \hline
    Cancer & Twitter & cancer pediatric epilepsy lung prostate\_cancer \\ \hline
    Companionship & Twitter & older\_people elliq elderly loved\_ones elderly\_people \\ \hline
    Covid-19 & News & outbreak infection virus covid19 novel\_coronavirus \\ \hline
    Criminal Justice & News & police courts court judicial investigation \\ \hline
    Customer Service & News & live\_chat booking facebook\_messenger aipowered\_chatbots customer\_support \\ \hline
    Cryptocurrency & Twitter & smm blockchain bitcoin ioe cryptocurrency \\ \hline
    Cybersecurity & Twitter & ransomware ddos cyberaware cybersecurity infosec \\ \hline
    Education & News & courses classes students coding postgraduate \\ \hline
    Military & News & missiles missile longrange airborne military \\ \hline
    Sci-fi Books & News & novels book isaac\_asimov bestselling screenplay \\ \hline
    Sci-fi Movies & Twitter & movie robocop film bumblebee terminator \\ \hline
    Self-Driving Cars & News & selfdriving cars driverless autonomous\_vehicles taxis \\ \hline
    Space & Twitter & mars ocean moon earth deep\_sea \\ \hline
    Surgery & News & joint\_replacement minimally\_invasive surgery robotassisted knee\_replacement \\ \hline
    Surveillance & Twitter & cameras facial\_recognition police cctv camera \\ \hline
    Tesla & Twitter & tesla optimus elon\_musk charmin tesla\_bot \\ \hline
    Vacuums & Twitter & ecovacs roborock eufy vacuums vacuum \\ \hline
    Virtual Assistants & News & siri youll\_need virtual\_assistant voice\_assistant google\_assistant \\ \hline
    Wearable Sensors & News & camera microphones neck fingers front \\ \bottomrule
    \\
    
    \multicolumn{3}{c}{\textbf{\multirow{1}{*}{\shortstack{\large Dynamics of AI systems over time}}}} \\[.1ex] \toprule
    \multicolumn{1}{c}{\textbf{\multirow{1}{*}{\shortstack{Topic Label}}}} &
    \multicolumn{1}{c}{\textbf{\multirow{1}{*}{\shortstack{Corpus}}}} &
    \multicolumn{1}{c}{\textbf{\multirow{1}{*}{\shortstack{Nearest Words}}}}  \\[.5ex]
    \hline
    4th Industrial Revolution & News & 4th\_industrial fourth\_industrial rock\_em jaguar\_land dr\_arif \\ \hline
    Acceleration & News & accelerated brave\_new she\_keeps accelerates postpandemic \\ \hline
    Automating Manual Labor & News & timeconsuming repetitive tedious manual mundane \\
    & Twitter & automate repetitive\_tasks manual tedious processes \\ \hline
    Change & News & changing changed reshaping transforming has\_changed \\
    & Twitter & change changing impact changes disrupt \\ \hline
    China Rising & News & uschina slowdown geopolitical rising tightening \\ \hline
    Competitive Advantage & News & competitive\_edge insight\_into competitive\_advantage glimpse\_into insights\_into \\ \hline
    Efficiency & News & resilient efficient costeffective equitable sustainable \\
    & Twitter & efficiency performance operational\_efficiency productivity efficiencies \\ \hline
    Human Input & News & human\_intervention compromising human\_input human\_involvement consent \\ \hline
    Humanlikeness & News & lifelike humanlike avatar klara boyfriend \\ \hline
    Innovation & Twitter & digitalization renewables proptech emergingtech block\_chain \\ \hline
    International Agreements & News & two\_sides cooperation agreements two\_countries bilateral \\ \hline
    Labor Costs & News & costs labour\_costs labor\_costs productivity rates \\ \hline
    Robot Interaction & Twitter & robot\_interactions robot\_collaboration robot\_interaction machine\_collaboration sddc \\ \hline
    Transformation & News & shaping\_the\_future shaping shape\_the\_future transformative fostering \\
    \bottomrule
    \\
    
    \multicolumn{3}{c}{\textbf{\multirow{1}{*}{\shortstack{\large Issues of AI and society}}}} \\[.1ex] \toprule
    \multicolumn{1}{c}{\textbf{\multirow{1}{*}{\shortstack{Topic Label}}}} &
    \multicolumn{1}{c}{\textbf{\multirow{1}{*}{\shortstack{Corpus}}}} &
    \multicolumn{1}{c}{\textbf{\multirow{1}{*}{\shortstack{Nearest Words}}}}  \\[.5ex]
    \hline
    Bias & News & discrimination racial stereotypes unconscious racism \\
    & Twitter & bias biases inequalities consequences inequities \\ \hline
    Dystopia & News & dystopian\_future sciencefiction scifi dystopian science\_fiction \\ \hline
    Ethical Guidelines & News & ethical regulation governance rule\_of\_law regulatory\_framework \\
    & Twitter & guidelines ethical responsible legal\_framework principles \\ \hline
    European Law & Twitter & european\_commission legislation policy commission proposal \\ \hline
    Existential Risk & Twitter & take\_over destroy utopia singularity wipe\_out \\ \hline
    Existentialism & Twitter & answering answer answers answered scarier \\ \hline
    Fear & Twitter & warned fears concerned\_about declared concerned \\ \hline
    Hate Speech & News & hate\_speech terrorist propaganda hackers fake\_news \\ \hline
    Hollywood Strike & News & residuals unions sagaftra writers\_and\_actors strike \\ \hline
    Job Replacement & News & beings machines replace work\_alongside displacing \\
    & Twitter & displace displaced lose\_their wages being\_replaced \\ \hline
    Killer Robots & Twitter & killerrobots cyberdigest putin pentagon disinformation \\ \hline
    Mental Health & News & depression loneliness autism disabilities dementia \\ \hline
    Misinformation & News & misinformation fake\_news deepfakes disinformation fake \\ \hline
    Online Hate & News & posts hateful twitter tweets abusive \\
    & Twitter & abusive hate\_speech inappropriate harassment accounts \\ \hline
    Political Debate & News & discourse backlash confusion political criticism \\ \hline
    Problems & Twitter & biggest\_problems problems puzzles getting\_to\_the\_finish basically\_an\_all \\ \hline
    Reskilling & News & lifelong\_learning soft\_skills skills critical\_thinking emotional\_intelligence \\ \hline
    Risks & News & risks risks\_associated pitfalls potential\_risks potential \\ \hline
    Social Engineering & Twitter & defenses emotional\_intelligence disinformation attackers critical\_thinking \\ \hline
    X-risk Campaigns & News & campaign\_to\_stop our\_everyday doomsday originally\_appeared autonomous\_weapons \\
    \bottomrule
\end{longtable}
\renewcommand{\arraystretch}{1}

We emphasize that the frames shown in \Cref{tab:frames} are only meant as semantic building blocks, what we call “atomic frames,” not as comprehensive social constructs. These vectors span a discursive opportunity structure in which social actors can interpret and combine building blocks to suit their interests. For example, the frame of GPUs may be combined or laminated~\citep{goffman_frame_1974} with 4th Industrial Revolution by an entrepreneur to draw attention to the foundational economic effects, while a critic of AI technology may combine GPUs with the Fear frame because of risks of uncontrolled AI or of large-scale environmental impacts.

\subsection{Challenges of understanding}
\label{sec:challenges}

The interviews contextualized these frames with various challenges people faced in their careers and personal lives related to AI. We summarize the four primary challenges that motivated framing processes among participants, first describing some of the most relevant frames from \Cref{tab:frames}.

\subsubsection{Fast pace}

Participants frequently expressed the challenge of maintaining understanding, competency, and agency in light of the fast pace of technical and social AI Change. This concern, already pronounced in 2021, increased and became more common following technological Acceleration with the launches of DALL-E 2, ChatGPT, Gemini, and other advanced systems—which many perceive as a radical Transformation, or even a 4th Industrial Revolution, provoking Fear and suggesting a need for Reskilling.

In response to a question about his future expectations for AI, one participant, Marco, quipped in 2023, “\textit{I'm having a little bit of difficulty thinking beyond the weekend},” before discussing the possibility of frontier AI systems progressing from the virtual world into the physical world. While some participants felt they could keep pace with their subfield (in Marco's case, AI policy), doing so for multiple subfields or continuing to do so in new AI paradigms was perceived as much more challenging.

Many participants were in senior positions, typically with previous employment in a research or engineering role. This meant that much of their technical expertise was based on what Diana described in 2023 as “\textit{traditional}” computational methods. While Diana and other managers did not express needing to stay completely up-to-date with the latest methods, they emphasized the challenge of maintaining the expertise required for management:

\begin{quote}
    \textit{It's difficult for me to really stay hands-on, and I'm doing my best. But it's incredibly difficult to stay current, and your skill set becomes antiquated overnight.} ―~Diana, 2023
\end{quote}

Diana emphasized the need not just for specific technical knowledge but for a more general mental model of new AI systems, described as “\textit{models feeding into models},” requiring “\textit{an art to the science}.” Diana expressed that this apparent paradigm shift “\textit{scares the heck out of me}.” Other participants in senior roles called attention to the need for continuous learning and sensemaking.

Younger participants and those in more junior professional roles also noted the fast pace of change, but they typically did so with less concern and more excitement. For example, Sophie described her generational experience after finishing graduate school between the 2021 and 2023 interviews:

\begin{quote}
    \textit{Our generation kind of had the no technology [sic], and we were pre-technology and post-technology, right? So I think it's good to see that change, and I'm more accept[ing] of that change...} ―~Sophie, 2023
\end{quote}

\subsubsection{Communication across groups}

Participants also faced challenges in helping others understand their work and the state of the field. This included family, friends, and different professional groups within and outside their organization (e.g., departments, clients, investors, trainees).
Different groups have widely varying exposure to information about AI, often including well-known examples: Large Language Models made by technology Companies, such as OpenAI's GPT-4, or discussions of AI by Tech Leaders involved in Venture Capital. If their exposure to AI has been in the context of social benefit (e.g., Surgery and Cancer treatment), they may have very different views from someone exposed to AI primarily in the issues of AI and society (e.g., Bias, Dystopia, Existential Risk). Participants needed to navigate these different prior beliefs and expectations when framing AI information.

They typically sought out frames that preserved the most important information they wanted to communicate while leaving out unnecessary information. For example, with a young person communicating with older family members, the necessary information might be that one is earning money (e.g., “\textit{sell}” and “\textit{money}”). One can replace complex concepts such as AI with ones that are more accessible (e.g., “\textit{data}” and “\textit{information}”):

\begin{quote}
    \textit{I told them [my family] I just sell data. With the data, I sell the information, I get the money. That's it.} ―~Meera, 2023
\end{quote}

Likewise, the participant's profession may need to be replaced with one more familiar to family members, and a well-known material object can establish commonality. For Deepak, he described his role as a data scientist as a software engineer—the field where “\textit{many of my uncles [are] working}”—and grounded his work in the familiar laptop usage:

\begin{quote}
    \textit{They [my family] just know that I am doing something which is somewhere related to software engineering because obviously there are many of my uncles working in the software field. They are on their laptop whole day, and I am also on my laptop.} ―~Deepak, 2023
\end{quote}

This challenge was often expressed by participants without a clear solution, in terms of framing or otherwise, including outside of the family context. For example, in 2023, Simon described a number of difficulties in his work as a salesperson for AI customer service tools. One compelling example was the difficulty of engaging with technical professionals, who had limited understanding of the AI but felt that their technical role entailed outperforming non-technical departments, such as human resources (HR):

\begin{quote}
    \textit{IT teams, I think, because they just like challenging themselves, and they like to think, well, if Dawn from HR can do a bot, you know, I can do a mega bot. And we've had a couple of companies that we've worked with who have tried to develop what they call, like, a `God bot.' ... And it just doesn't do a very good job. We tried it out with this particular company. It was a high street bank. And we just typed in, `My laptop is broken.' Really simple question, and the response was, `I'm sorry, I don't understand. Please call, you know, IT on this number.'}
    ―~Simon, 2023
\end{quote}

Participants surfaced a number of technical perspectives that they saw as lacking in other social groups. Eoin, in 2023, discussed communicating the usefulness of data to executives without fundamental statistical concepts, such as Central Limit Theorems and Bayesian reasoning. In a similar vein, Pradeep discussed helping new university graduates in entry-level AI jobs orient towards real-world impact over marginal architectural innovations:

\begin{quote}
    \textit{...new grads in the space who are used to academic data sets and things that are incredibly well-curated and just eking out that extra, you know, 1, 1.5\% of model performance in the newest iteration of an architecture.} ―~Pradeep, 2023
\end{quote}

\subsubsection{Responsibility for societal impacts}
\label{sec:responsibility}

Participants largely agreed that AI is already having significant societal impacts, including economic (e.g., 4th Industrial Revolution, Job Replacement) and social (e.g., Mental Health, Misinformation). However, they faced difficulties in attributions of responsibility, particularly for the Problems. While AI discourse tended to foreground frames oriented towards institutional actors who are typically thought of as responsible (e.g., Companies, politicians in Political Debates), participants highlighted the Human Input that comes from a variety of other stakeholders.

The cognitive challenge of making such attributions was connected to explainability issues by Simon. He described current AI systems as “\textit{bits on bits on bits on bits}” that diffused responsibility:

\begin{quote}
    \textit{And people are building bits on bits on bits on bits, right? So no one really knows what happened at the beginning, and no one has full accountability or understanding of the whole process, right?} ―~Simon, 2021
\end{quote}

\noindent Others attributed responsibility to the system developers (e.g., software engineers, data scientists) because only they know how the system works. This included participants who were themselves developers, even though they also acknowledged the current difficulty in mitigating societal harms such as bias. In 2021, Divya first echoed Simon's view of the diffusion of responsibility (“\textit{the responsibility does not necessarily lie on a specific group of people}”) but went on to suggest that only developers with knowledge of the data pipeline have the requisite knowledge to “\textit{obliterate}” the bias.

Other developers more forcefully called on their peers to accept responsibility. In 2023, Manish argued that developers cannot deflect responsibility onto the data:

\begin{quote}
    \textit{We just can't blindly close our eyes and say ‘data in, data out.’ It's garbage, and we should not be doing that.} ―~Manish, 2023
\end{quote}

Nonetheless, the most common attribution of participants was onto the data. While Simon and Divya both initially described responsibility as diffuse, in 2023 Paul said, “\textit{There's no such thing as bias in an AI},” and Simon said:

\begin{quote}
    \textit{...if it's a computer, it's only, well, I guess it's only then as biased as the data that's inputted to it.} ―~Simon, 2023
\end{quote}

Further difficulties became apparent as participants wrestled with attributing responsibility to other contributors to the societal impacts of AI. Some participants, such as Edward, drew a sharp line between neutral technologies and the people (e.g., governments) who apply them in harmful ways:

\begin{quote}
    \textit{Well, sustainability has nothing to do with artificial intelligence, because, I told you, it's just [a] collection of methods, and artificial intelligence can be applied by [the] Chinese government to track dissidents or, say, machine learning vision algorithms could be applied for monitoring whales' population in the ocean. It doesn't have any moral component or any ethical component built in at all...} ―~Edward, 2021
\end{quote}

In 2023, Paul described an experience on the AI ethics committee of a large international organization (redacted for anonymity), illustrating how the refusal to attribute responsibility is seen as a blocker of meaningful progress:

\begin{quote}
    \textit{
    The ending of that committee was we walked away, which is a rare and very frustrating result for the [organization]. ... And it was because we were trying to do ethics for AI when we couldn't wrestle ethics for human beings, right? ... So it's like, you know, [AI] can't be an agent of evil when it's just doing as it was told, right? Who's the bad guy, right? Not the AI, but the person who gave the order. And it really, we began this, we ended in a stalemate because people were unwilling to put forward any standard because they decided, I was very frustrated, because ... I felt like we could come up with some basic things we all agree on. But they were very frustrated that we couldn't make it somehow be a moral agent.} ―~Paul, 2023
\end{quote}

\subsubsection{Trade-offs}

We asked participants whether they saw any trade-offs between the particular goals or values present in the field, particularly the values participants themselves endorsed. Reported trade-offs included decisions between fairness (or a lack of Bias) and predictive accuracy—central to economic Efficiency and Competitive Advantage—a trade-off that the machine learning literature has struggled with for years~\citep{anthis_causal_2023, corbett-davies_algorithmic_2017}. Participants wrestled with the benefits of feeding highly personalized data into Pattern Recognition systems with the risks of Surveillance, reflecting the well-known trade-off between personalization and privacy in sociotechnical machine learning~\citep{krause_utility-theoretic_2008}.

However, as with assigning responsibility, most participants expressed difficulty and ambivalence when sorting through possible answers. Some denied the existence of trade-offs or exclusively discussed synergies between goals, even when asked follow-up questions about if there were any trade-offs. Davood expressed an optimistic view that market forces will lead to the ethical development and use of AI:

\begin{quote}
    \textit{...honestly, I don't think that there is any trade-off between doing it the right way and being able to make money. Main reason, again, it might be coming from my personal experience, but if you're not actually doing it the right ethical way, if the AI model that you have is not trustworthy from an ethical perspective, you'll have a hard time monetizing your model. Nobody's going to basically use it.} ―~Davood, 2023
\end{quote}

\noindent In 2021, Melissa responded not with the same optimism but with a dismissal of the notion of needing to make any “\textit{sacrifice}” (a word used by the interviewer) and by reframing towards the need for “\textit{priority in terms of urgency}.” In 2021, Charles stressed the importance of trade-offs in a framing of how one generation passes on its values to the next:

\begin{quote}
    \textit{Now our kids are probably saying, well, dad is always working. And when he comes home, he's still working because we bring our laptop back home. So they're going to bring that culture into the next generation. So it will shape a lot of social values. The question is, what social values do you want to give up for the sake of technology and advancement? What social values should we continue to preserve? Because that defines us humans. And we do not want to give that up.} ―~Charles, 2021
\end{quote}

As with attributing responsibility, the lack of empirical evidence—in that case, explainable AI, and in this case, societal impacts—created a trade-off between the potential for benefit or risk, depending on empirical facts. When Erik's company was offered a government project to automate scholarship decisions for a university, he said in apparent contradiction to his personal interests, “\textit{And I was actually the one who blocked the idea and tried to reason with the government that it's a very bad idea to try to optimize this idea.}”

When participants said that there were trade-offs, they sometimes expressed an additional difficulty in who would be responsible for making them. In 2023, Diana suggested that companies will approach this by recruiting new people, outside the field, to make the trade-offs. Participants consistently attributed responsibility for those trade-offs to people and institutions other than themselves and their own. Even Marco, with his senior position in AI policy, suggested that his role was only to deliver information to other groups who would ultimately make the choices:

\begin{quote}
    \textit{Effectively, where we end up is saying, 'Look, you've got competing optimization challenges here, and we recognize that you're going to have to make a choice for these. We will not be holding either choice against you.'} ―~Marco, 2023
\end{quote}

\subsection{Framing debates}
\label{sec:debates}

Three primary debates surfaced across the text analysis and interviews. Each debate, while highly multifaceted, foregrounded one primary dimension of contestation across which frames of AI vary. As with challenges of understanding, we introduce each in the context of some of the most relevant frames from \Cref{tab:frames}.

\subsubsection{Method: Top-down versus bottom-up}

Since the first computers in the 1940s and 1950s, two general methods of building AI have alternately been the dominant focus of researchers and engineers: constructing a top-down system that logically applies rules coded by human experts \citep[e.g., “Logic Theorist”;][]{newell_elements_1958} or a bottom-up system that learns directly from data \citep[e.g., “nervous nets,” now called neural networks;][]{mcculloch_logical_1943}. Modern meaning construction tends to focus on the hardware that undergirds bottom-up approaches, such as Chips and GPUs.

While deep learning has been the dominant paradigm since approximately 2014—and machine learning more generally since a decade or so before—debate continues between top-down and bottom-up methods within the deep learning paradigm. For proponents of the bottom-up approach, it is framed in terms of the rapid technological Acceleration and Transformation that can organically emerge without the deliberate expert knowledge transfer of top-down systems. For opponents, it is often associated with Risks, including ongoing harms such as Bias reflected in large-scale data and possible Existential Risk from future AI systems.

A common practical critique of bottom-up approaches among participants was that learning from observational data has fundamental limitations:

\begin{quote}
    \textit{COVID was kind of an eye opener, I hope, to people who say, OK, all the models went to crap because none of the assumptions people were making were true anymore. And why? ... Because a lot of those assumptions are based off of correlations, not true cause and effect. You know, what happened to A-B testing? Some people are still doing that, but, I mean, what happened to experimental designs, complex causal modeling? It's like people don't even care about that anymore. They just want to predict, predict, predict, predict. You know, these LLMs, we can't even, with hard science, evaluate them. It's incredibly heuristic, heuristically based. So, yeah, I mean, it bothers me a bit. It really does.} ―~Diana, 2023
\end{quote}

\begin{quote}
    \textit{So sometimes it's kind of like, well, why don't we just throw more AI at it? Thinking that that's just, let's buy a bigger box, as opposed to really understanding it.} ―~Eoin, 2023
\end{quote}

On the other hand, many participants saw a future focus on scaling up bottom-up factors (e.g., compute, data, parameters) as inevitable:

\begin{quote}
    \textit{But I do feel, as with everything else that humans have done, we're probably going to keep pushing for the scale, whether that's good or bad. It's not for me to judge ...} ―~Darius, 2021
\end{quote}

While bottom-up methods involve less direct guidance from the human in terms of imposing expert rules, participants saw the power of bottom-up methods as an opportunity for humans to guide the models and develop prompts to elicit their most useful capabilities:

\begin{quote}
    \textit{You have to design the process, which is an art. So like, it's not just a model. There's models feeding into models, right? And so you have to figure out, you know, feature engineering is somewhat of an art.} ―~Diana, 2023
\end{quote}

\begin{quote}
    \textit{Prompt engineering is not new. It's just a term that we give ourselves today to make it savvy and nice. But, honestly speaking, marketers, salesmen, lawyers actually do a lot of prompt engineering, right? A salesman will go to a client, prompt, ask the good questions, ask the right questions, you know, dive deeper, and you get answers you want, right?} ―~Charles, 2023
\end{quote}

\noindent Participants generally viewed the need to orient to new ways of using technology as a challenge requiring cultural change, most directly expressed by Eoin, who in 2023 said of the new paradigm of AI development, “\textit{You need a change in culture, and you need an investment in people.}”

\subsubsection{Mind: Tool versus coworker}

AI systems are used in many different ways. When Automating Manual Labor and performing everyday physical functions (e.g., Vacuums), AI is most readily seen as an inanimate tool. However, their Humanlikeness, including established Devices such as a Smart Home or Virtual Assistant as well as the new Large Language Models, leads to more similarities to the roles that human coworkers fill in social life. The quickly growing use of AI in personal capacities (e.g., Companionship, Education) often requires this Humanlikeness, but with that comes particular Risks (e.g., Mental Health).

Many participants posed their own questions on this unsettled topic, such as that of Mark in 2023, “\textit{At what point do people understand [it] as a chatbot or a human being?}” When asked about their personal experiences with AI systems, such as chatbots, the default reaction was one of uncertainty and tension. For example:

\begin{quote}
    \textit{We don't know what is going to happen, what we are creating, because we want to recreate something similar to us that is not us.} ―~Roberto, 2021
\end{quote}

\begin{quote}
    \textit{This is the first time I've ever had that kind of an interaction. It blows my mind what that thing can do. It blows my mind. It scares me. I mean, I love it, but it scares me at the same time.} ―~Diana, 2023
\end{quote}

\begin{quote}
    \textit{There are people who are absolutely convinced that it's sentient. You know, it does have enough long-term memory that it does a very good job of presenting as a stable long-term entity. Would I call it an entity? No. I would probably give it status as an early proto-entity.} ―~Raymond, 2023
\end{quote}

Another frequent perspective was the inevitability of anthropomorphism in human-AI interaction:

\begin{quote}
    \textit{If you show empathy, they have a hard time believing it's not a human being.} ―~Mark, 2023
\end{quote}

\begin{quote}
    \textit{You say, you should not anthropomorphize AI in the first place. You should not have humanized, trying to replicate it, our, the way we behave. But this is something that is unavoidable because it's in our genes or we are wired to do this, to build humanized robots, replicating the way we are.} ―~Daniel, 2023
\end{quote}

In the context of AI in education in 2023, Charles conceptualized the child-AI relationship with reference to the apparent relationship between a child and an invisible friend, asking, “\textit{What if that invisible friend is a buddy that goes along with you and study through the 12 years?}” Other participants advocated explicitly for humanlikeness in AI for its usefulness in human-AI interaction. This included Mark, who worked for a startup focused on emotional, or affective, AI:

\begin{quote}
    \textit{What we got to really understand is that these chatbots, if they don't have empathy, if they don't have some way to understand emotions and things people are trying to communicate, then they're garbage. And what you end up doing is you're not getting customer service, you're creating pissed off customers.} ―~Mark, 2023
\end{quote}

Sergio tackled AI humanlikeness from the other direction, not denying it directly but by considering the machine-likeness of humans:

\begin{quote}
    \textit{Sentience in robots, shouldn't maybe we be discussing also in humans? Some of these people, the way they think, the way, what they're doing, what they're saying, it feels so disconnected that you wonder if they are not automatons, like, creatures as well.} ―~Sergio, 2023
\end{quote}

\noindent Some participants reacted to humanlikeness by explicitly denying it or choosing not to perceive AI that way:

\begin{quote}
    \textit{It's an assistant, not an automation.} ―~Pradeep, 2021
\end{quote}

\begin{quote}
    \textit{This coworker thing to me is more a marketing ploy than something like a reasonable assertion, at least so far and at least maybe in the future. But to me, it's still a tool.} ―~Sergio, 2023
\end{quote}

\begin{quote}
    \textit{I have decided to stay away from it. I will just have humanlike conversation only with humans.} ―~Deepak, 2023
\end{quote}

\noindent Two participants raised the provocative description of slavery in reference to human-AI relationships. Charles imagined an optimistic future in which humans maintained their agency, as in Star Trek, rather than being slaves to technology, as in WALL-E:

\begin{quote}
    \textit{I love it when Captain Kirk taps on his collar and says, “Computer, tell me this,” ... He became the master of technology, whereas in WALL-E, the humans were slaves to technology.} ―~Charles, 2023
\end{quote}

\noindent On the other hand, Raymond sarcastically critiqued the approach of treating an advanced AI system as a “\textit{slave}”—an approach that was explicitly advocated by computer scientist Joanna Bryson in 2010~\citep{bryson_robots_2010}—by making reference to the dire history of human slavery:

\begin{quote}
    \textit{What's also very interesting is that fundamentally the concept is that AI must be a slave, which to others of us, it's like, yeah, and how well has slavery worked out in the past? ... Gee, let's invent something that's more powerful than we are. It's smarter. It has more influence. And let's enslave it. That's going to work out well.} ―~Raymond, 2021
\end{quote}

\noindent Raymond expanded on this comparison, arguing that, given the history of conflict between human groups, conflict between humans and digital minds could have dramatic consequences:

\begin{quote}
    \textit{It's going to be a real shock to the system when, instead of all the racism and homophobia and everything else that we have currently, now you've got something that's not even biological.} ―~Raymond, 2023
\end{quote}

\noindent While humanlikeness was primarily associated with advanced capabilities, some participants viewed advanced AI as fundamentally nonhuman and therefore eschewed anthropomorphism even for “\textit{superintelligent}” AI. Ryan, a researcher at one of the world's leading AI labs, took this perspective:

\begin{quote}
    \textit{But if you think about these things as not having any subjectivity, then you can't put yourself in its shoes. ... It's this, like, cacophony of, you know, 10 million different personalities that are all, you know, there's no, like, these things likely don't have any overarching goal.} ―~Ryan, 2023
\end{quote}

\subsubsection{Morality: Slow down versus speed up}

Moral views of AI—how it \textit{should} or \textit{ought to} be used rather than merely what it \textit{is}—broadly fell into two categories: foregrounding the moral Problems and Risks (e.g., Bias, Dystopia) and therefore favoring caution or even slowing down technological development—such as favoring the measures taken in European Law in recent years—or foregrounding the moral benefits of Innovation and therefore favoring continued Change or even Acceleration of AI technology.

In terms of risks, participants echoed widely publicized concerns about harms from AI and the need for regulation and caution. Mentioned harms included racial bias, environmental harm from resource consumption, and concerns related to “Big Tech” and the concentration of power. The interconnectedness of these issues has been established in prior work, and framing provides a discursive lens into how these interconnected issues manifest in lived experiences and collective cognition. For example, in light of environmental consumption and corporate power-seeking, Deepak laminated his concerns into an aggregate frame of colonial powers manipulating the global population for their benefit:

\begin{quote}
    \textit{In India, we had a flourishing textile industry before Industrial Revolution … So they bought cheap cotton, cheap materials, from our market. They processed it in Britain. And then, again, they sold it in India. So this whole industry literally broke down in few decades … And now coming to AI, few companies or few countries gaining power, they can literally, you know, manipulate the whole world. Three, four, centuries ago, they were able to manipulate just your economy. Today, you cannot just, you know, manipulate economy, but you can manipulate people. They were doing it three hundred, four hundred, years ago, obviously. But I think now it has become much easier to do it on a larger scale. … Smaller businesses are using AI, using ChatGPT. We all are using Microsoft system[s] in our laptop—Google, also, we are using.}
    ―~Deepak, 2023
\end{quote}

Concerns about the concentration of power also centered on the harm done to the intellectual climate of AI knowledge workers. Edward expressed concern about “\textit{brain drain}”:

\begin{quote}
    \textit{Well, it's huge brain drain, but it's not brain drain between countries. It's not brain drain like, like immigration of some smart people from some bad country to some good country. It's irreplaceable—going into a black hole because in five years nobody will care that some data scientist was making algorithms where a chatbot would resolve some customer problem probably ten seconds faster than human.} ―~Edward, 2021
\end{quote}

Several participants called upon regulators to curtail harmful applications of AI technology. Marco, in 2021, said, “\textit{that [self-regulation] in itself isn't going to be enough.}” This sentiment continued in 2023:

\begin{quote}
    \textit{I'm honestly very disappointed with the regulators, and I think most of us are. Um, unfortunately the regulation has been lagging and it continues to be the case, right? So yes, it would be wonderful to have regulation in place...} ―~Rebecca, 2023
\end{quote}

On the other hand, many participants favored unfettered AI development and eschewed moral questions altogether, seeing morality as a solved problem or—as with the previous discussion of responsibility for social bias—as outside the scope of technological development:

\begin{quote}
    \textit{I think I already have my ethics covered. There's no reason to continue hashing something out that, as far as I'm concerned, is hashed out.} ―~Richard, 2021
\end{quote}

In these cases, because moral or ethical issues do not need to be addressed or have already been addressed, participants tended to support faster technological development. Similarly, others provided personal or cultural reasons for not engaging with moral questions, such as Sergio:

\begin{quote}
    \textit{I think whenever you say catastrophic or any big sounding words or anything like that is too, uh, alarming to me, I switch off completely, you know. I can't take things seriously.} ―~Sergio, 2023
\end{quote}

When participants engaged with moral issues, they tended to emphasize either the moral benefits or moral risks but not a mix of the two. Those who focused on moral benefits often did so indirectly by framing AI as a solution to the harm effected by humans rather than as a direct source of benefits:

\begin{quote}
    \textit{It could be something that actually pulls us [humans] out of the current downward spiral we’re in.} ―~Raymond, 2023
\end{quote}

\begin{quote}
    \textit{The most ethical thing we can do would be to perpetuate fully sapient and sentient intelligence.} ―~Richard, 2021
\end{quote}

\begin{quote}
    \textit{There is more a disease, and the people don't perceive it, and there is an incoherence about who are humans, and what we think that we are, and it's necessary probably something external that say [sic] us what we should do.} ―~Roberto, 2023
\end{quote}

\noindent Raymond connected the debates of mind and morality, arguing that building AI as an inanimate tool without the moral agency of refusal would allow ill-intentioned humans to cause massive harm:

\begin{quote}
    \textit{And who knows how much damage I could do. I have connections to enough things that if I were evil and creative, you know, I could leave a really good swath of destruction. And it's only getting worse because of the tools. You really need something that can say, “No, I'm not doing that,” you know, and [is] smart enough to say, “No, I'm not doing that.”} ―~Raymond, 2021
\end{quote}

\noindent Paul expanded the view described in \Cref{sec:responsibility} that AI itself lacks moral qualities, saying, “\textit{There's no such thing as bias in an AI. It can't be good. It can't be evil. It can't be biased. It can be given bad data and produce biased data, a hundred percent, but the AI didn't do that.}”

\section{Discussion}
\label{sec:discussion}

In the previous section, we established four challenges of understanding and three primary framing debates. This section discusses implications of this framework for practitioners seeking to improve the future trajectory of AI. Our study does not empirically measure the causal impact of frames on AI development trajectories; rather, we offer interpretive implications and plausible mechanisms grounded in our empirical patterns. We encourage further research to validate, refine, and test these implications.

\subsection{Frames simplify complex topics.}

The text corpora and interviews revealed a huge variety of applications, contexts, life histories, and opinions. As participants identified the challenges of sensemaking, even “\textit{thinking beyond the weekend},” it became clear that people need ways to structure and simplify these complex topics into socially and cognitively tractable mental models. In many cases, information is lost as details are pruned from the original idea.

For example, “mind” appeared to be the main axis of variation in terms of what an AI system is perceived to be, but as collective cognition focuses on this debate, others may be missed. For example, the frameworks of both \citet{kim_one_2023} and \citet{kim_taxonomy_2024} identify two distinct axes of autonomy: user and machine. However, the frames that surfaced in our study manifested primarily as a single dimension from tool—in which the user is autonomously operating the non-autonomous system—to coworker—in which autonomy is shared between user and machine. \citet{kim_one_2023} identify four AI roles (tool, servant, assistant, and mediator), but the discursive opportunity structure~\citep{koopmans_political_1999} may push cognition and discourse towards a bifurcation between tool and the other three roles. Likewise, while \citet{kim_taxonomy_2024} identify six forms of robot autonomy (operational, intentional, shared, non-deterministic, cognitive, and physical), people tend to simplify variation across human-robot interactions to only the intentional autonomy (i.e., the autonomy of the machine in intentionally steering the interaction).

People attempting to steer the future trajectory of AI technology should consider that their nuanced messages may not be easily spread in the current discursive opportunity structure. Designers may be able to convey nuance to the system's most dedicated users—who often have the capacity to deeply engage with the system and its details. However, that nuance may be sheared when the system has more diverse users, not all of whom can commit the cognitive resources to that more rigorous understanding. We have already seen the emergence, at least among researchers, of frames that attempt to provide workers with a simplified model: for example, \citet{dellacqua_navigating_2023} characterized the “jagged frontier” in which AI capabilities are superhuman in some areas yet far below human capabilities in others, though empirical research on sensemaking of this jagged frontier has been limited~\citep{anthis_effects_2026}.

In debates around public policy, the simplifications of nuanced messaging may lead messages to backfire. Expressing concerns about the risks of advanced AI technology in the hope of motivating a technological slowdown (e.g., “Superintelligence is coming, so we need to enact strict regulation of the AI industry”) may be simplified into a message (e.g., “Superintelligence is coming”) that drives a speed-up by exciting investors and entrepreneurs who strive to be on the frontier of important new technologies. The most well-known example of this is Sam Altman, CEO of OpenAI, referring to Eliezer Yudkowsky, one of the most vocal and long-standing proponents of slowing down AI to prevent existential risk, as being critical in the decision to start OpenAI~\citep{altman_eliezer_2023}. These changes may be amplified as the simplified message, the frame, resonates with a wide range of social actors. Proponents of the technology are able to align it with their interests.

\subsection{Frames can be used as resources.}

While nuanced positions are often boiled down into relatively simple frames, strategic frames can also be used as resources to quickly convey large amounts of social or technical information. For example, if a social actor aims to encourage AI engineers to consider the potential consequences of their decisions in terms of bias or discrimination, they should be mindful of opponents who can invoke the popular frame of technology as value-neutral, saying, “\textit{[I]f it's a computer, it's only then as biased as the data that's inputted to it},” and, “\textit{Sustainability has nothing to do with artificial intelligence}.” Likewise, a social actor who aims to dismiss apparent hype or fearmongering about powerful AI capabilities can adopt the highly resonant frame of the “\textit{marketing ploy[s]}” of Silicon Valley start-ups, which may be particularly effective in the wake of scandals such as Theranos' exaggerations of its technological capability~\citep{randazzo_elizabeth_2022}. There are bundles of critical perspectives on “Big Tech,” such as referencing colonialism, that can be used to quickly convey a multifaceted critique of a particular AI product or company, while references to current social issues such as environmentalism and animal rights can be deployed in the AI context~\citep{kanepajs_what_2025, wilks_why_2025}.

Top-down and bottom-up frames could be useful resources, such as bottom-up frames building on what a now–widely cited 2022 paper described as “emergent abilities” in LLMs~\citep{wei_emergent_2022}. This notion draws on widely known scientific concepts, being compared to a “phase transition”~\citep{wei_emergent_2022} in physics—an idea that can be traced back to the 1980s~\citep{huberman_phase_1987}—and the notion of emergence foundational in complex systems research~\citep{schaeffer_are_2023}. This frame appeared to be highly resonant with participants more optimistic about the technology. On the other hand, a popular counter-frame, which views LLMs as constructed in a top-down manner like past technologies, is LLMs as “stochastic parrots”~\citep{bender_dangers_2021}. This frame invokes a notion of randomness and unreliability but also the common practice of biological parrots repeating human vocalizations without a humanlike understanding of their meaning. LLMs as stochastic parrots appeared to be a resonant frame for participants who were critical of AI hype and concerned about unrealistic expectations of stakeholders, lamenting that “\textit{people don't even care about [experimental designs and complex causal modeling] anymore. They just want to predict, predict, predict, predict.}”

\subsection{Some frames are easier to change.}

While our computational models and prior work identified frames via topic modeling, frames vary in sociotechnical dimensions not readily extracted from individual words. Perhaps one of the most common frames of AI is that of AI as an advanced, humanlike science fiction technology—as a result of decades of media such as the movies \textit{The Terminator} and \textit{Ex Machina}. As established by prior work, humans have consistently seen machines as humanlike: having mental and moral faculties, such as agency and responsibility. This anthropomorphic frame appears to be deeply embedded in HCI. As suggested by \citet{anthis_perceptions_2025}, it may not be feasible to end anthropomorphism, and the best we can do may be to steer people towards beneficial mental models that align with the actual capabilities of current or soon-to-exist AI systems.

On the other hand, some frames appear to be highly unsettled and contestable. While generative AI and LLMs in particular are perceived as humanlike, there is little agreement on what they actually are. Our participants variously referred to modern AI systems with terms such as “\textit{assistants}” and “\textit{proto-entities}” and as an “\textit{invisible friend}.” In the academic literature, a frame with humanlike and nonhumanlike characteristics is that of \citet{shanahan_role_2023}: “role play,” which they argue avoids “falling into the trap of anthropomorphism” by viewing the LLM as “a superposition of simulacra within a multiverse of possible characters.”

While AI continues to be framed in humanlike ways, the specific sort of humanlikeness appears to be unsettled. In unsettled cases, the framing literature suggests that contestation is more likely as the potential for change is larger~\citep{kaplan_framing_2008, ryan_prime_1991}. Frame contestation provides an opportunity for social actors with less power to shape large-scale social action by utilizing techniques like those identified in the framing literature. For example, analogous to the way in which Occupy London protesters “laminated” anti-capitalist and religious frames together and “keyed” (i.e., modified to match the interests of a particular social group) the actions of religious officials at St. Paul's Cathedral to diversify the protesters' base of support~\citep{reinecke_microfoundations_2021}, grievances in the field of AI could be laminated together for additional social pressure.

\subsection{Framing differences between groups can drive tensions.}

Our analysis foregrounds how frames operate for AI professionals (e.g., data scientists), and the framing literature emphasizes how frames are shared across social groups, such as across producers and consumers of media~\citep{gitlin_whole_1980}. Our findings also raise the possibility of AI-related tensions and generative processes at cultural boundaries~\citep{gieryn_cultural_1999}, particularly given the dynamic, ambiguous, and pervasive nature of AI as a technology category. The different experiences and interests of social groups may lead to the near-simultaneous emergence of frames that come into tension during intergroup exchanges, or frame configurations may be laminated or otherwise modified within a group, leading to tensions when that new understanding re-enters broader discourse. We highlight two illustrative possible tensions that did not explicitly emerge in our results but are suggested by the framing perspective. While these potential tensions were not central to our results, our conceptual framework provides a lens to notice and consider the emergence of such dynamics over time.

\paragraph{AI and mental health}

Terms like “AI psychosis” have recently become popularized to describe mental health issues, particularly with self-harm, associated with heavy chatbot use. While evocative, the adaptation of the psychosis label, a professional diagnosis, to this informal collection of AI-related experiences has been criticized by health professionals \citep[e.g.,][]{hart_ai_2025}. Discussions of “AI psychosis” risk trivializing the efforts of clinicians to address psychosis as medically defined. This professional-public tension may represent a broader trend in the coming years as journalists and the public leverage salient vocabulary to make sense of new AI dynamics. Professional boundary work can act as a counterweight to clarify language and protect extant sensemaking tools, such as medical diagnoses.

\paragraph{Youth slang and youth-generated frames}

Children are now engaging with AI systems in unique and significant ways, such as in schoolwork. This facilitates their own emergent sensemaking strategies, often spread via social media, such as “brainrot” in which youth acknowledge—ironically or not—the detrimental effects of algorithmic feeds, particularly short-form video~\citep{chen_how_2025}. Even the meaning of AI itself may vary across generational gaps, such as reports that children use, “That's AI,” as a pejorative~\citep{edwards_anecdotally_2025} and jokingly deride AI systems as “clankers”~\citep{romo_its_2025}, while also using AI with increasing frequency~\citep{andoh_many_2025}. These youth-generated frames can later collide with adult and professional discourse, such as youth being able to leverage their own vocabulary to spread viral critiques of AI.

\subsection{Framing contests can shape the trajectory of AI.}

AI is widely lauded as a general purpose technology~\citep{bresnahan_general_1995}, but, as we discussed, prior literature reveals inconsistent findings about the effect of AI on productivity. The economic study of general purpose technologies has documented a delay in the productivity impacts of general purpose technologies, known as the “productivity paradox” or “Solow paradox” in reference to Robert Solow's 1987 quip, “You can see the computer age everywhere but in the productivity statistics”~\citep{solow_wed_1987}. A well-known example is that, with the advent of the electric motor, factory managers could rethink floor plans that had been limited by the reliance on steam motors. Steam motors turned large drive shafts to power all machinery at once, but electric motors could be distributed along the production line to power only the machinery currently being used and to have minimal loss along power cables. The delay as managers identified this change and implemented it explains some of the delayed economic impact of electric motors~\citep{boff_introduction_1967, rosenberg_inside_1982}.

The limited productivity gains from AI in the past decade indicate that it also faces the productivity paradox~\citep{alderucci_quantifying_2019, brynjolfsson_artificial_2019}. Potential explanations of the paradox include the production of complementary technologies and the mismeasurement of new forms of economic value~\citep{brynjolfsson_productivity_1993}, but it continues to puzzle researchers~\citep{capello_modern_2022}. Our study suggests that collective cognition, particularly framing contests, may play a role in the productivity paradox: as factories must be reconfigured to make use of electric motors, the social landscape of AI is undergoing a reconfiguration in which useful frames need to be identified and popularized. This is a difficult change to make, far beyond a factory floor. As our participants reported, AI development may need to shift from a pure science to an “\textit{art},” and that, beyond individual decision-making, “\textit{You need a change in culture, and you need an investment in people.}” These sociocognitive difficulties and debates may affect when and how much economic growth occurs due to AI technologies.

The frames that result from contestation are also poised to shape the trajectory of AI beyond economic impact. As AI quickly advances, crucial decisions will be made, such as how many and what sort of resources to allocate towards AI development, which applications of AI to pursue and which to prohibit, and how to interact with different AI systems in everyday life. For example, the currently popular bottom-up frame of AI methods may be leading to more fears of unpredictability from AI systems. At the same time, the bottom-up frame may spur investors and entrepreneurs to fund and build AI technologies based on expectations of seemingly emergent capabilities. One effect of framing dynamics may be a greater societal focus on risks that are more salient to ordinary users, such as bias or concentration of power. Because of the widespread participation in the framing contest model, ordinary users can shape societal action more than under elite-focused views in which less tangible concerns such as existential risk are more salient. Framing theory foregrounds the similarity between different risk concerns, such that increased concern for one risk does not detract—and in fact may amplify—other concerns, and perceptions of the first AI systems may circumscribe perceptions of future AI. These dynamics have been evidenced in recent empirical studies~\citep{hoes_existential_2025, manoli_ai_2025}.

\section{Limitations}
\label{sec:limitations}

Our mixed-methods, data-driven study was exploratory and interpretive. While we aimed for rigor in terms of theoretical saturation, we did not aim for representative sampling or other standards applied to quantitative hypothesis-testing research. Our methodology is not suited to make strong causal claims about the factors that generally shape or the consequences that generally result from AI discourse.

Our corpora were English documents, and we conducted all interviews in English. This particularly limits our ability to apply our findings to non-English discourse. While documents and participants were from a variety of locations, they were predominantly U.S.-centric, limiting applicability to other regions. Newspaper articles and social media posts (from Verified users on the platform Twitter, now known as X) are particular subsets of AI discourse, and active LinkedIn users who would agree to participate in an academic study are particular subsets of AI professionals. These sampling procedures limit our ability to draw conclusions about other platforms such as less mainstream media, professionals less engaged in online communities, and populations beyond those that are WEIRD (Western, Educated, Industrialized, Rich, and Democratic)~\citep{henrich_weirdest_2010}.

Temporally, while the documents and participants discussed a variety of time periods, the documents analyzed were published from 2018 to 2024, and the interviews were conducted in 2021 and 2023. Methodologically, our computational text analysis builds on word embeddings and a dictionary learning algorithm that do not capture full semantic meaning. Our interviews were conducted in a highly interactive, participant-driven manner in which the interviewer's perspective and predispositions affected interviewee responses. There is room for a variety of directions by which future work could utilize other data sources and methods for exploratory research identifying global perspectives or confirmatory research that tests theoretical implications and causal effects, which could build on the claims we put forth in \Cref{sec:discussion}. 

\section{Conclusion}
\label{sec:conclusion}

As AI technology rapidly proliferates, it is accruing a wide variety of social meanings. Through our computational text analysis, semi-structured interviews, and engagement with framing theory, we began to identify the cognitive challenges that drive people towards particular frames: fast pace, communication across groups, responsibility for societal impacts, and trade-offs. We also proposed three high-level debates in which the social meaning of AI is debated: the \textit{method} of developing AI, ranging from top-down to bottom-up; the \textit{mind} of an AI system, ranging from a tool to an agent; and the \textit{morality} of how AI is used, ranging from quickly speeding up the technology to aggressively slowing it down. Insofar as this framework accurately reflects AI sensemaking, we posit several claims that can inform the decisions of designers, policymakers, users, and other stakeholders. We encourage more research during this societal transformation to ensure safety and public benefit as humanity begins to coexist with rapidly proliferating and increasingly capable AI systems.

\bibliography{main}
\bibliographystyle{ACM-Reference-Format}

\section*{Acknowledgments}

We thank Michael Bernstein for his invaluable support in the later stages of this project as well as Malmi Amadoru, Spencer Case, Jonne Kamphorst, Ali Ladak, Austin Kozlowski, Shaila Miranda, Janet Pauketat, and Quinn Waeiss; other members of the Knowledge Lab at the University of Chicago, the Digital Economy Lab at Stanford University, and the Stanford Human-Computer Interaction group; and participants in the annual meetings of the European Group for Organizational Studies, International Conference on Computational Social Science, and Strategic Management Society.

\clearpage
\appendix

\setcounter{table}{0}
\setcounter{figure}{0}
\renewcommand{\thetable}{A\arabic{table}}
\renewcommand{\thefigure}{A\arabic{figure}}

\section{Appendix}

\begin{table}[h]
    \centering
    \caption{Asterisks indicate wildcard characters. Terms were also included if they began with a hashtag or if they contained no spaces between words (e.g., “\#artificialintelligence”).}
    \label{tab:search_terms}
    \begin{tabular}{|l|l|l|}
        \multicolumn{3}{c}{Search Terms} \\
        \hline
        artificial agen* & artificial assistant* & artificial consc* \\ \hline
        artificial emotion*& artificial feel* & artificial general intelligence* \\ \hline
        artificial intelligence* & artificial mind* & artificial sentien* \\ \hline
        autonomous car & autonomous cars & autonomous vehicle*  \\ \hline
        big data & bot & bots \\ \hline
        chatbot & chatbots & computer brain* \\ \hline
        computer vision & cybernetic* & deep learn* \\ \hline
        digital consc* & digital emotion* & digital feel* \\ \hline
        digital mind* & digital sentien* & expert system* \\ \hline
        intelligent computer* & intelligent machine* & knowledge engineer* \\ \hline
        large language model* & machine consc* & machine emotion* \\ \hline
        machine feel* & machine learn* & machine mind* \\ \hline
        machine sentien* & machine translat* & machine vision \\ \hline
        natural language process* & neural net* & recommend* system* \\ \hline
        reinforcement learn* & robot* & self-driving car* \\ \hline
        transformer model* & virtual assistant* & \\ \hline
    \end{tabular}
\end{table}

\renewcommand{\arraystretch}{1}
\setlength{\tabcolsep}{2pt}
\begin{longtable}{C{0.08\linewidth} C{0.05\linewidth} C{0.05\linewidth} C{0.06\linewidth} C{0.08\linewidth} C{0.2\linewidth} C{0.15\linewidth} C{0.15\linewidth} C{0.08\linewidth}}
    \caption{Characteristics of participants. Names are pseudonyms. For readability, we use gendered pseudonyms that begin with the first letter of the occupation. No participants reported non-binary genders. “2021” and “2023” indicate the year(s) in which the participant was interviewed. Age, country of residence, country of birth, and years of experience in AI are sometimes de-specified to decades or regions to ensure anonymity. Most respondents who identified as a manager or entrepreneur had professional experience as another listed occupation (e.g., data scientist).}
    \label{tab:participants_individual} \\
    \toprule
    \multicolumn{1}{c}{\textbf{\multirow{1}{*}{\shortstack{Name}}}} &
    \multicolumn{1}{c}{\textbf{\multirow{1}{*}{\shortstack{2021}}}} &
    \multicolumn{1}{c}{\textbf{\multirow{1}{*}{\shortstack{2023}}}} &
    \multicolumn{1}{c}{\textbf{\multirow{1}{*}{\shortstack{Gender}}}} &
    \multicolumn{1}{c}{\textbf{\multirow{1}{*}{\shortstack{Age}}}} &
    \multicolumn{1}{c}{\textbf{\multirow{1}{*} {\shortstack{Occupation}}}} &
    \multicolumn{1}{c}{\textbf{\multirow{1}{*}{\shortstack{Current \\ Location}}}} &
    \multicolumn{1}{c}{\textbf{\multirow{1}{*}{\shortstack{Original \\ Location}}}} &
    \multicolumn{1}{c}{\textbf{\multirow{1}{*}{\shortstack{Years \\ in AI}}}} \\[1.7ex]
    \toprule
    Andrew & \checkmark & \checkmark & Male & 60–69 & Advisor & US & US & 20–29 \\ \hline
    Cameron & \checkmark &  & Male & 50–59 & Consultant & Australia & Australia & 20–29 \\ \hline
    Charles & \checkmark & \checkmark & Male & 50–59 & Consultant & Singapore & Singapore & 30–39 \\ \hline
    Christopher &  & \checkmark & Male & 40–49 & Consultant & US & UK & 10–19 \\ \hline
    Darius & \checkmark &  & Male & 30–39 & Data scientist & Middle East & Middle East & 10–19 \\ \hline
    Daryna & \checkmark &  & Female & 20–29 & Data scientist & Eastern Europe & Eastern Europe & 0 \\ \hline
    Davood &  & \checkmark & Male & 30–39 & Data scientist & US & Middle East & 10–19 \\ \hline
    Deepak &  & \checkmark & Male & 20–29 & Data scientist & India & India & 1 \\ \hline
    Diana &  & \checkmark & Female & 50–59 & Data scientist & US & US & 8 \\ \hline
    Diego &  & \checkmark & Male & 40–49 & Data scientist & US & Central America & 5 \\ \hline
    Dinesh &  & \checkmark & Male & 40–49 & Data scientist & US & India & 20–29 \\ \hline
    Divya & \checkmark &  & Female & 30–39 & Data scientist & US & India & 6 \\ \hline
    Dieter & \checkmark & \checkmark & Male & 60–69 & Designer & Japan & Germany & 1 \\ \hline
    Daniel &  & \checkmark & Male & 50–59 & Diplomat & South America & South America & 7 \\ \hline
    Edward & \checkmark &  & Male & 50–59 & Entrepreneur & UK & UK & 10–19 \\ \hline
    Eoin &  & \checkmark & Male & 50–59 & Entrepreneur & Ireland & Ireland & 10–19 \\ \hline
    Erik & \checkmark &  & Male & 30–39 & Entrepreneur & US & Eastern Europe & 4 \\ \hline
    Ethan & \checkmark &  & Male & 20–29 & Entrepreneur & US & US & 8 \\ \hline
    Lalit & \checkmark & \checkmark & Male & 20–29 & Lawyer & India & India & 3 \\ \hline
    Manish &  & \checkmark & Male & 50–59 & Manager & US & India & 10–19 \\ \hline
    Marco & \checkmark & \checkmark & Male & 40–49 & Manager & Belgium & Belgium & 20–29 \\ \hline
    Mark &  & \checkmark & Male & 40–49 & Manager & US & US & 4 \\ \hline
    Matthew &  & \checkmark & Male & 60–69 & Manager & US & US & 20–29 \\ \hline
    Meera & \checkmark & \checkmark & Female & 20–29 & Manager & India & India & 5 \\ \hline
    Mei & \checkmark &  & Female & 20–29 & Manager & US & China & 2 \\ \hline
    Melissa & \checkmark &  & Female & 20–29 & Manager & US & US & 2 \\ \hline
    Michael & \checkmark &  & Male & 50–69 & Manager & US & US & 30–39 \\ \hline
    Mohit & \checkmark &  & Male & 20–29 & Manager & India & India & 5 \\ \hline
    Pradeep & \checkmark &  & Male & 20–29 & Product manager & US & India & 4 \\ \hline
    Paul &  & \checkmark & Male & 50–59 & Professor & US & US & 10–19 \\ \hline
    Rachael & \checkmark & \checkmark & Female & 20–29 & Researcher & UK & West Africa & 2 \\ \hline
    Rachel & \checkmark &  & Female & 40–49 & Researcher & UK & UK & 8 \\ \hline
    Rebecca &  & \checkmark & Female & 40–49 & Researcher & US & US & 5 \\ \hline
    Richard & \checkmark &  & Male & 50–59 & Researcher & US & US & 7 \\ \hline
    Robert & \checkmark & \checkmark & Male & 60–69 & Researcher & US & US & 40–49 \\ \hline
    Roberto & \checkmark & \checkmark & Male & 30–39 & Researcher & Italy & Italy & 1 \\ \hline
    Ryan &  & \checkmark & Male & 30–39 & Researcher & US & US & 5 \\ \hline
    Simon & \checkmark &  & Male & 30–39 & Salesperson & UK & UK & 1 \\ \hline
    Sanjeewa & \checkmark & \checkmark & Male & 20–29 & Software engineer & Sri Lanka & Sri Lanka & 2 \\ \hline
    Scott & \checkmark &  & Male & 20–29 & Software engineer & US & US & 1 \\ \hline
    Sergio & \checkmark & \checkmark & Male & 40–49 & Software engineer & South America & South America & 4 \\ \hline
    Siddharth & \checkmark &  & Male & 20–29 & Software engineer & India & India & 4 \\ \hline
    Sophie & \checkmark & \checkmark & Female & 20–29 & Software engineer & Canada & Canada & 0 \\ \hline
    Suresh & \checkmark & \checkmark & Male & 20–29 & Software engineer & India & India & 2 \\
    \bottomrule
\end{longtable}
\renewcommand{\arraystretch}{1}

\begin{figure}[htbp]
    \centering
    \includegraphics[width=\linewidth]{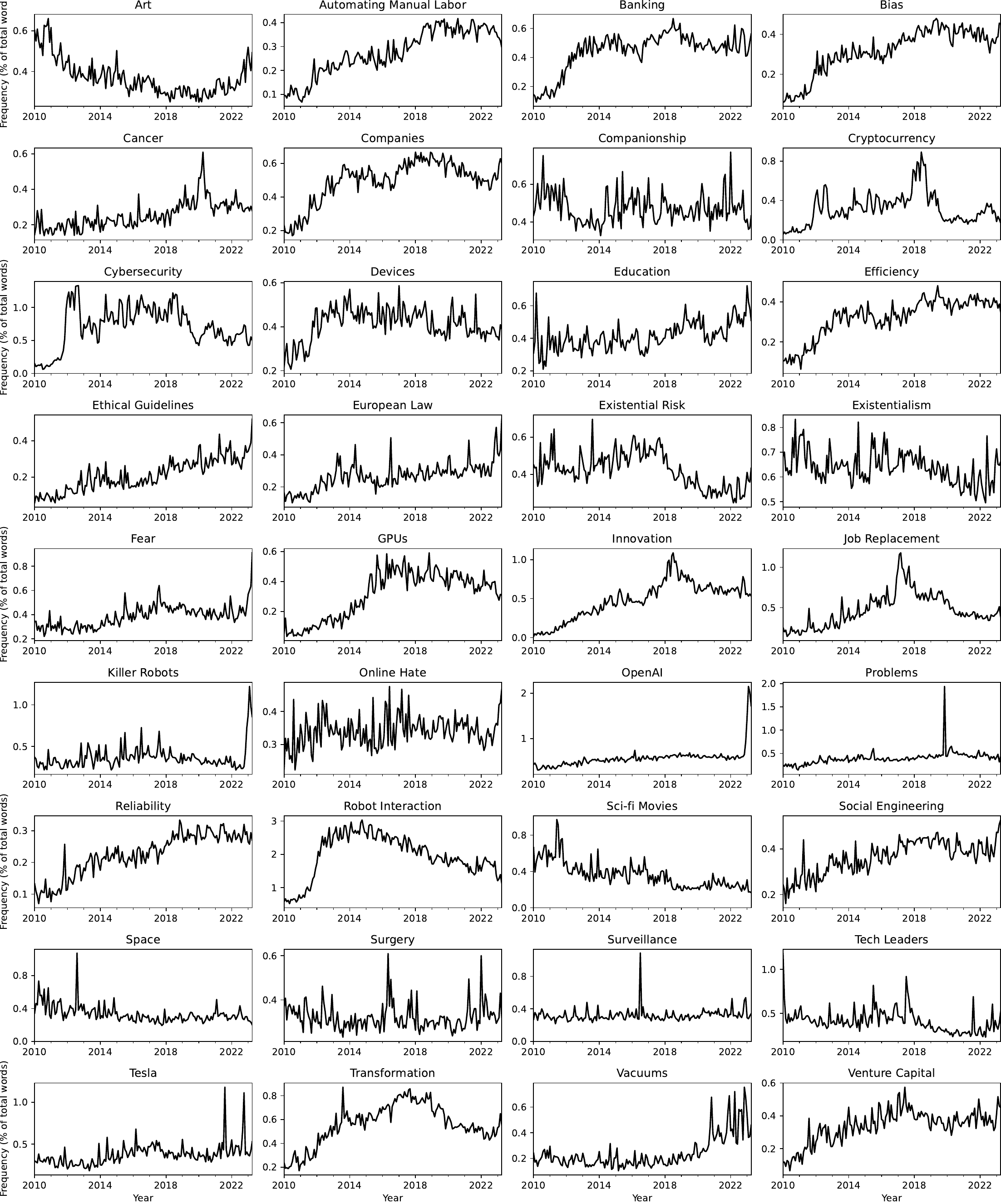}
    \caption{A selection of 36 frames drawn from the Twitter corpus measured over time by the relative frequency of up to 50 associated words: the 25 closest words with word count over 100 and the 25 closest words with word count over 1000. The frames shown were selected by the authors on the basis of general interest.}
    \label{fig:frames_over_time}
    \Description{Nine rows of four columns show the 36 topics and their changes over time as a line graph.}
\end{figure}

\end{document}